\documentclass[a4paper,12pt]{article}
\usepackage[utf8]{inputenc}
\usepackage{eucal, amssymb,amsmath,graphicx, amsfonts, latexsym,url, subfigure, hhline,lineno}
\usepackage[font={small,it}]{caption}
\usepackage{setspace}
\newcommand{\ind}{1\hspace{-2.5mm}{1}}
\title{Inferring $R_0$ in emerging epidemics --\\ the effect of common population structure is small}
\date{}
\author{Pieter Trapman,$^{1\ast}$ Frank Ball,$^{2}$ Jean-St{\'e}phane Dhersin,$^{3}$ \\ Viet Chi Tran,$^{4}$ Jacco Wallinga,$^{5,6}$ and Tom Britton$^{1}$\\
\\
\normalsize{$^{1}$Department of Mathematics, Stockholm University, Sweden}\\
\normalsize{$^{2}$School of Mathematical Sciences, University of Nottingham, UK}\\
\normalsize{$^{3}$LAGA, CNRS (UMR 7539), Universit{\'e} Paris 13, Sorbonne Paris Cit{\'e},  France}\\
\normalsize{$^{4}$Laboratoire Paul Painlev{\'e}, Universit{\'e} des Sciences et Technologies de Lille, France}\\
\normalsize{$^{5}$Rijksinstituut voor Volksgezondheid en Milieu (RIVM), Bilthoven, The Netherlands}\\
\normalsize{$^{6}$Department of Medical Statistics and Bioinformatics,}\\
\normalsize{Leiden University Medical Center, Leiden, The Netherlands}\\
\normalsize{$^*$ Corresponding author, email: \url{ptrapman@math.su.se} }
}

\begin{document}
 \maketitle
\begin{abstract}

When controlling an emerging outbreak of an infectious disease it is essential to know the key epidemiological parameters, such as the basic reproduction number $R_0$ and the control effort required to prevent a large outbreak. These parameters are estimated from the observed incidence of new cases and information about the infectious contact structures of the population in which the disease spreads. However, the relevant infectious contact structures for new, emerging infections are often unknown or hard to obtain. Here we show that for many common true underlying heterogeneous contact structures, the simplification to neglect such structures and instead assume that all contacts are made homogeneously in the whole population, results in conservative estimates for $R_0$ and the required control effort. This means that robust control policies can be planned during the early stages of an outbreak, using such conservative estimates of the required control effort.
\end{abstract}

\baselineskip16pt

\section{Introduction}

An important area of infectious disease epidemiology is concerned with the planning for mitigation and control of new emerging epidemics. The importance of such planning has been highlighted during epidemics over recent decades, such as HIV around 1980  \ \cite{Fauc08}, SARS in 2002/2003 \ \cite{BLCG05}, A H1N1 influenza pandemic in 2009 \ \cite{Y09} and the Ebola outbreak in West Africa, which started in 2014  \cite{teamebola}. A key priority is the early and rapid assessment of the transmission potential of the emerging infection. This transmission potential is often summarized by the expected number of new infections caused by a typical infected individual during the early phase of the outbreak, and is usually denoted by the basic reproduction number, $R_0$. Another key priority is estimation of the proportion of infected individuals we should isolate before they become infectious in order to break the chain of transmission. This quantity is denoted as the required control effort $v_c$. If a fully efficient vaccine is available, the required control effort is equal to the proportion of the population that needs to be vaccinated in order to stop the outbreak, if the people receiving the vaccine are chosen uniformly at random. These key quantities are inferred from available observations on symptom onset dates of cases and the generation times, i.e., the typical duration between time of infection of a case and infection of its infector \cite{WL07,SSAG10}.  The inference procedure for $R_0$ and $v_c$ requires information on the infectious contact structure (``who contacts whom''), information that is typically not available or hard to obtain quickly for emerging infections.

The novelty of this paper lies in that we assess estimators for the basic reproduction number $R_0$ and required control effort $v_c$, which are based on usually available observations,  over a wide range of assumptions about the underlying infectious contact structure. 
We find that most plausible contact structures result in only slightly different estimates of $R_0$ and $v_c$. Furthermore, we find that ignoring the infectious contact pattern, thus effectively assuming that individuals mix homogeneously, will in many cases result in a slight \textit{overestimation} of these key epidemiological quantities, even if the actual contact structure is far from homogeneous. This is important good news for planning for mitigation and control of  emerging infections, since the relevant contact structure is typically unknown: ignoring the contact structures results in slightly conservative estimates for $R_0$ and $v_c$. This is a significant justification for basing infection control policies on estimates of $R_0$ derived for the Ebola outbreak in West Africa in \cite{teamebola}, where the data are stratified by region, without further assumptions on contact structure.

We focus on communicable diseases that follow an infection cycle where the end of the infectious period is followed by long-lasting immunity or death. In such an infection cycle, individuals are either susceptible, exposed (latently infected), infectious or removed (which means either recovered and permanently immune or dead).  Those dynamics can be described by the so-called stochastic SEIR epidemic model \cite[Ch.3]{DHB12}.  For ease of presentation we use  the Markov SIR epidemic as a leading example. In this special case, there is no latent period (so an individual is able to infect other individuals as soon as they are infected), the infectious period is exponentially distributed with expected length $1/\gamma$,  and infected individuals make close contacts at a constant rate $\lambda$. 
While infectious, an individual infects all susceptible individuals with whom he or she has close contact. The rate at which an infectious individual makes contact with other individuals depends on the contact structure in the community but it does not change over time in the Markov SIR model. The more general results for the full SEIR epidemic model are given and derived in Appendix A. 

\begin{figure*}
\centering
\subfigure[ ]{
    \includegraphics[width=.13\textwidth]{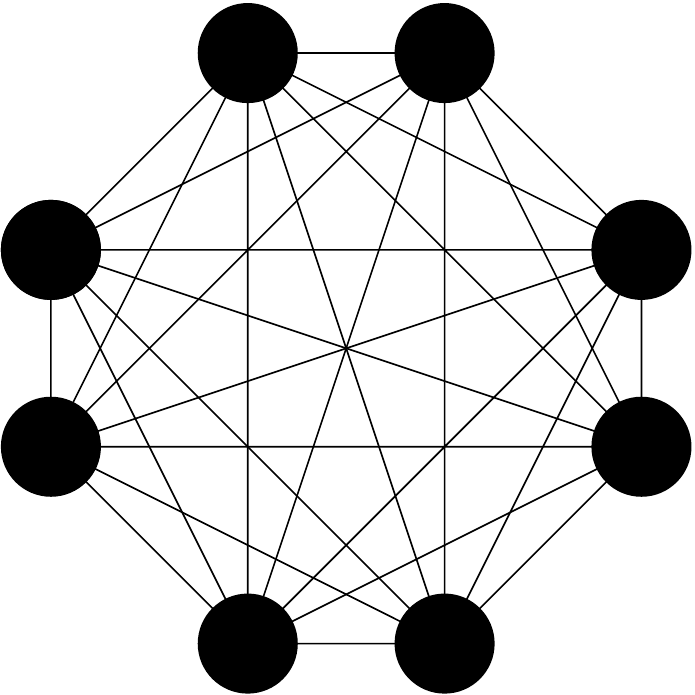}
}
\hspace{1cm}
\subfigure[ ]{
   \includegraphics[width=.13\textwidth]{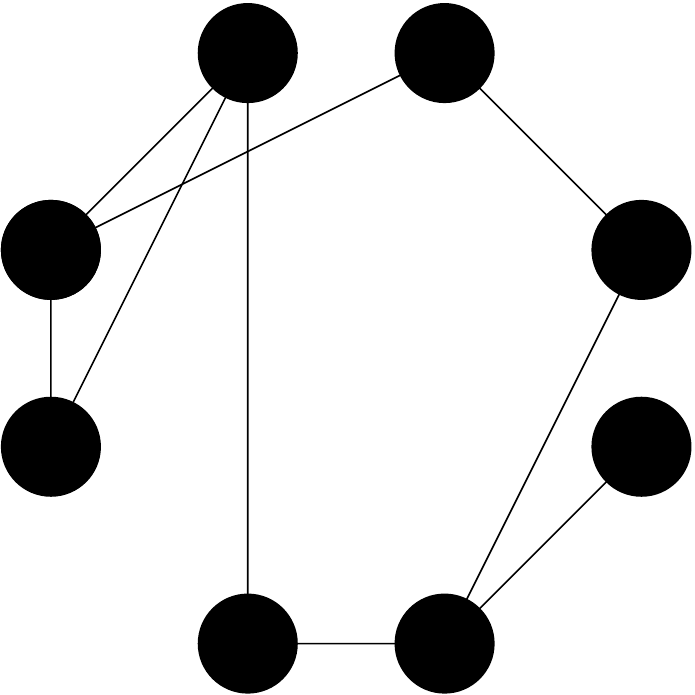}
}
\hspace{1cm}
\subfigure[ ]{
   \includegraphics[width=.13\textwidth]{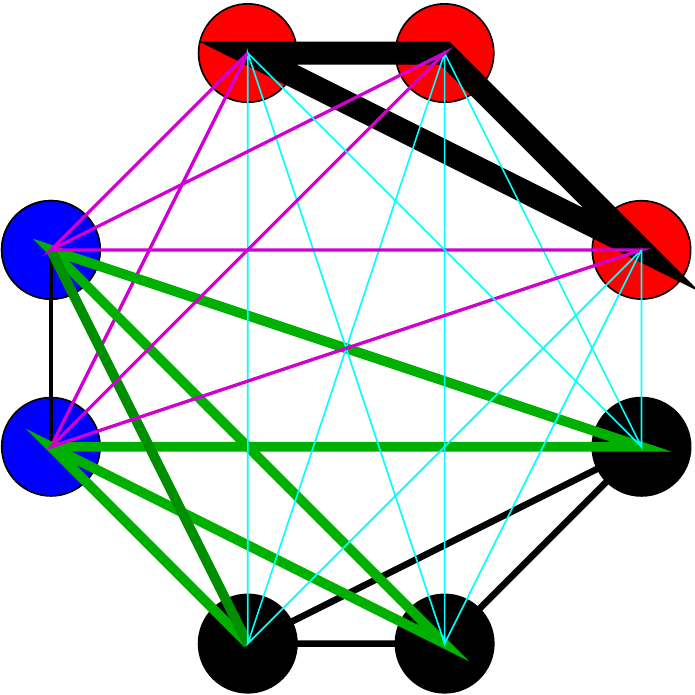}
}
\hspace{1cm}
\subfigure[ ]{
    \includegraphics[width=.13\textwidth]{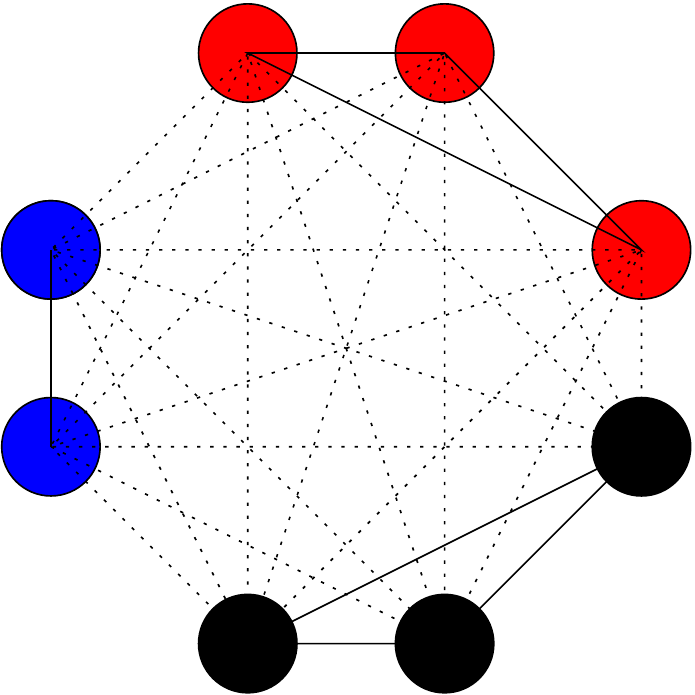}
}
\caption{The four contact structures considered: individuals are represented by circles and possible contacts are denoted by lines between them. (a) A homogeneous mixing population, in which all individuals have the same frequency of contacting each other. (b) A network structured population, in which, if contact between two individuals is possible, the contacts occur at the same frequency. (c) A multi-type structure with three types of individuals, in which individuals of the same type have the same colour and lines of different colour and width represent different contact frequencies. (d) A population partitioned into 3 households, in which members of the same households have the same colour and household contacts, represented by solid lines, have higher frequency than global contacts, represented by dotted lines.  }\label{popstructs}
\end{figure*}

We cover a wide range of possible contact structures. For each of these we derive estimators of the basic reproduction number and the required control effort. We start with the absence of structure, when the individuals mix homogeneously \cite[Ch.1]{AM91} (Figure \ref{popstructs}(a)).  We examine three different kinds of heterogeneities in contacts: the first kind, network structure \cite{Ande98,Barb13,DDMT12,Newm02} (Figure \ref{popstructs}b), emphasizes that individuals have regular contacts with only a limited number of other individuals; the second kind, multi-type structure (Figure \ref{popstructs}c), emphasizes that individuals can be categorized into different types, such as age classes, where differences in contact behaviour with respect to disease transmission are pronounced among individuals of different type but negligible among individuals of the same type  \cite{BC93,DHB12}; and the third kind, household structure \cite{Beck95,Ball97} (Figure \ref{popstructs}d),  emphasizes that individuals tend to make most contacts in small social circles, such as households, school classes or workplaces. Finally, we compare the performance of the estimators for $R_0$ and $v_c$ against the simulated spread of an epidemic on an empirical contact network.

\section{Estimation of $R_0$ and required control efforts for various contact structures}

\subsection{Homogeneous mixing}
Many results for epidemics in large homogeneous mixing populations can be obtained since the initial phase of the epidemic is well approximated by a branching process \cite{Ball95,Jage75,Hacc05}, for which an extensive body of theory is available. In particular, an outbreak can become large only if $R_0>1$. Note that if $R_0>1$, then it is still possible that the epidemic will go extinct quickly. The probability for this to happen can be computed \cite[Eq.\ 3.10]{DHB12} and is less than 1. Another result is that if $R_0>1$  and the epidemic grows large (which we assume from now on), then the number of infectious individuals grows roughly proportional to $e^{\alpha t}$ during the initial phase of the epidemic. Here $t$ is the time since the start of the epidemic and the epidemic growth rate $\alpha$ is a positive constant, which depends on the parameters of the model, through the equation
\begin{equation}
\label{malthus}
1= \int_0^{\infty} e^{-\alpha t} \beta(t) dt.
\end{equation}
Here $\beta(t)$ is the expected rate at which an infected individual infects other individuals $t$ time units after they were infected. For the Markov SIR model, with expected duration of the infectious period $1/\gamma$, $\beta(t)$ is given by  $\lambda e^{-\gamma t}$.
This can be understood by observing that $\lambda$ is the rate at which an infected individual makes contacts if he or she is still infectious, while $e^{-\gamma t}$ is the probability that the individual is still infectious $t$ time units after he or she became infected.
The epidemic growth rate $\alpha$ corresponds to the Malthusian parameter for population growth. 
Note that the expected number of newly infected individuals caused by a given infected individual equals
\begin{equation}
\label{R0def}
R_0 = \int_0^{\infty} \beta(t) dt.
\end{equation}
For the Markov SIR model, (\ref{malthus}) and (\ref{R0def}) translate to 
\begin{equation}
 \label{Homcom}
 1 = \frac{\lambda}{\gamma + \alpha} \qquad \mbox{and} \qquad R_0 = \frac{\lambda}{\gamma}.  
\end{equation}

Since we usually have observations on symptom onset dates of cases for a new, emerging epidemic, as was the case for the Ebola epidemic in West Africa,  it is often possible to estimate $\alpha$ from observations. In addition, we often have observations on the typical duration between time of infection of a case and infection of its infector, which allow us to estimate, assuming a Markov SIR model, the average duration of the infectious period, $1/\gamma$ \cite{WL07}. 
Using (\ref{Homcom}), this provides us with an estimator of $R_0$ in a homogeneously mixing Markov SIR model:
\begin{equation}
\label{R0alpharel}
R_0 =1+\frac{\alpha}{\gamma},
\end{equation}
which, as desired, does not depend on $\lambda$.
In Appendix A we deduce expressions for $\alpha$ and $R_0$, in terms of the model parameters for the more general SEIR epidemic and relate those quantities.

The required control effort for the SEIR epidemic in a homogeneously mixing population is known to depend solely on $R_0$ through the relation \cite[p.69]{DHB12} 
\begin{equation}
\label{vstandrel}
v_c = 1- \frac{1}{R_0}.
\end{equation}
 Thus, we obtain an estimator of the required control effort in terms of observable growth rate and duration of infectious period: 
\begin{equation}
\label{valpharel}
v_c = \frac{\alpha}{\alpha + \gamma}.
 \end{equation}
We compare the estimators (\ref{R0alpharel}) and (\ref{valpharel}) with other estimators that we obtain for different infectious contact structures, using the same values for the epidemic growth rate and duration of the infectious period. Throughout the comparison we assume that the initial stage of an epidemic shows exponential growth, which is a reasonable assumption for many  diseases, including the Ebola epidemic in West Africa. 

\subsection{Network structure}\label{sec:netw}
One kind of infectious contact structure is network structure. We consider the so-called configuration model \cite{Newm03},\cite[Ch.3]{Durr06} in which each individual may contact only a limited number (which varies between individuals) of other acquaintances, with mean $\mu$ and variance $\sigma^2$.
In such a network, the mean number of  different individuals (acquaintances) a typical newly infected individual can contact (other than his or her infector) is referred to as the mean excess degree \cite{Newm03}, which is given by \[\kappa = \frac{\sigma^2}{\mu} +\mu-1\] (see Appendix A.4 or \cite{Newm03} for the derivation of $\kappa$). 
This quantity is hard to observe for a new emerging infection, but we know the value must be finite and strictly greater than 1 if the epidemic grows exponentially fast. For the Markov SIR model for which the constant rate at which close contacts per pair of acquaintances occur is denoted by $\lambda^{(net)}$, we obtain $\beta(t) = \kappa \lambda^{(net)} e^{-(\lambda^{(net)} + \gamma) t}$. This can be seen by noting that $\kappa$ is the expected number of susceptible acquaintances a typical newly infected individual has in the early stages of the epidemic, while $e^{-\lambda^{(net)}t}$ is the probability that a given susceptible individual is not contacted by the infective over a period of $t$ time units, and $e^{-\gamma t}$ is the probability that the infectious individual is still infectious $t$ time units after he or she became infected.
In Appendix A we deduce an estimator of $R_0$ in terms of the observable epidemic growth rate, the average duration of the infectious period and the unobservable mean excess degree: $R_0=\frac{\gamma + \alpha}{\gamma + \alpha/\kappa}$ (c.f.\ \cite{Pell15}). 
We find that the estimator obtained assuming homogeneous mixing (\ref{R0alpharel}) overestimates $R_0$ by a factor 
$1+\frac{\alpha}{\gamma \kappa}.$

We know that this factor is strictly greater than 1, since the exponential growth rate $\alpha$, the recovery rate $\gamma$ and mean excess degree $\kappa$ (which is often hard to observe) are all strictly positive.

In Appendix A we also consider more general SEIR models.
We conclude that estimates of $R_0$ obtained by assuming homogeneous mixing are always larger than the corresponding estimates if the contact structure follows the configuration network model. In Appendix A.4.5 we show by example, that if we allow for even more general random infection cycle profiles, then it is possible that assuming homogeneous mixing might lead to a non-conservative estimate of $R_0$. However, for virtually all standard models studied in the literature, assuming homogeneous mixing leads to conservative estimates. 

As is the case for the homogeneous mixing contact structure, the required control effort for epidemics on the network structures under consideration, is known to depend solely on $R_0$ through equation (\ref{vstandrel}) \cite{Brit07}. This provides us with an estimator of $v_c$ in terms of observable $\alpha$ and duration of infectious period and the unobservable mean excess degree $\kappa$: 
$v_c = \frac{\kappa -1}{\kappa}\frac{\alpha}{\alpha + \gamma}$.
We find that the estimator obtained assuming homogeneous mixing overestimates $v_c$ by a factor 
$1+ \frac{1}{\kappa-1}.$
This factor is always strictly greater than 1, since the mean excess degree $\kappa$ is strictly greater than 1. Thus, $v_c$ obtained by assuming homogeneous mixing is always larger than that of the configuration network model. Consequently we conclude that, if the actual infectious contact structure is made up of a configuration network and a perfect vaccine is available, we need to vaccinate a smaller proportion of the population than predicted assuming homogeneous mixing. 

The overestimation of $R_0$ is small whenever $R_0$ is not much larger than 1 or when $\kappa$ is large. The same conclusion applies to the required control effort $v_c$.
The observation that the $R_0$ and $v_c$ for the homogeneous mixing model exceed the corresponding values for the network model extends to the full epidemic model allowing for an arbitrarily distributed latent period followed by an arbitrarily distributed independent infectious period, during which the infectivity profile (the rate of close contacts) may vary over time but depends only on the time since the start of the infectious period (see Appendix A.4.4 for the corresponding equations). Figure \ref{ratiosfig}a  shows that for SIR epidemics with Gamma distributed infectious periods, the factor by which the homogeneous mixing estimator overestimates the actual $R_0$ increases with increasing epidemic growth rate $\alpha$, and suggests that this factor increases with increasing standard deviation of the infectious period. Figure \ref{ratiosfig}b shows that the factors by which the homogeneous mixing estimator overestimates the actual $v_c$, decreases with increasing $\alpha$ and increases with increasing standard deviation of the infectious period. When the standard deviation of the infectious period is low, which is a realistic assumption for most emerging infectious diseases (see e.g.\ \cite{Cori12}), and $R_0$ is not much larger than 1, then ignoring the contact structure in the network model and using the simpler estimators for the homogeneous mixing results in a slight overestimation of $R_0$ and $v_c$.

\begin{figure}
\centering
\subfigure[ ]{
    \includegraphics[width=.4\textwidth]{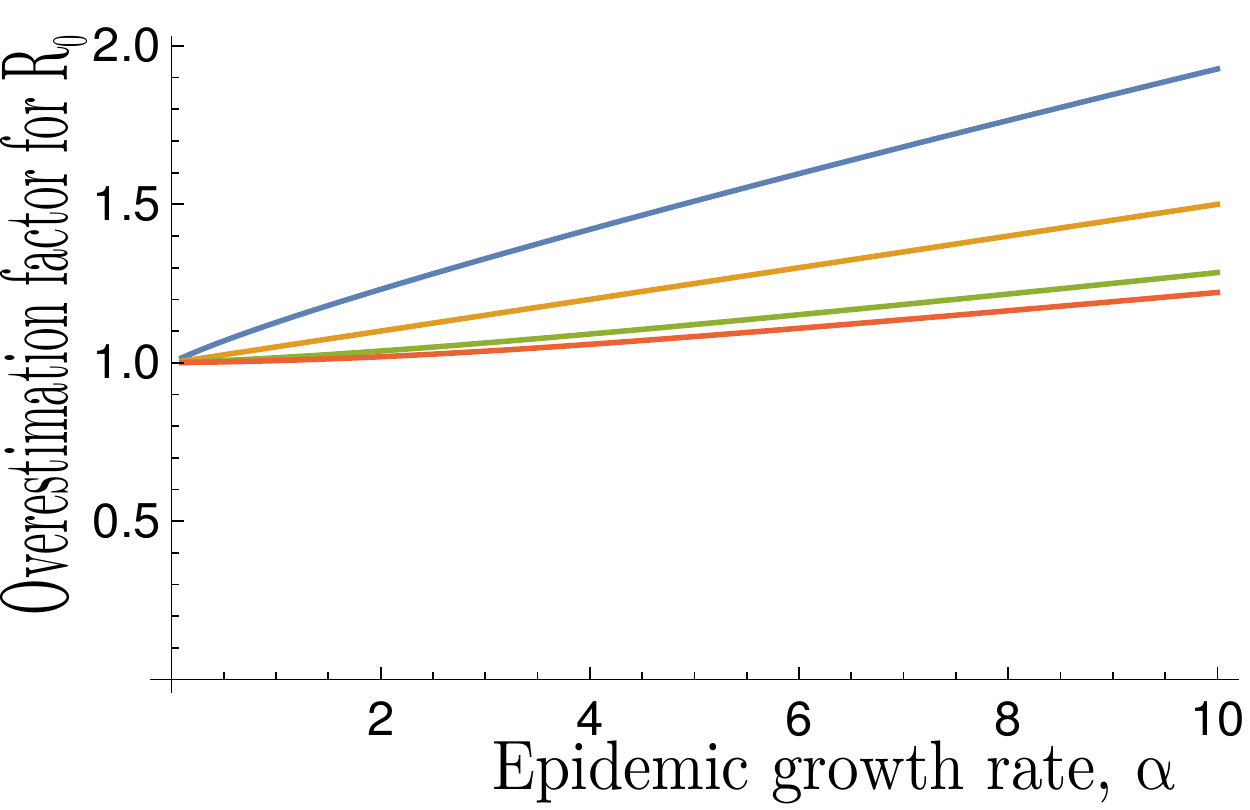}
}
\subfigure[ ]{
   \includegraphics[width=.4\textwidth]{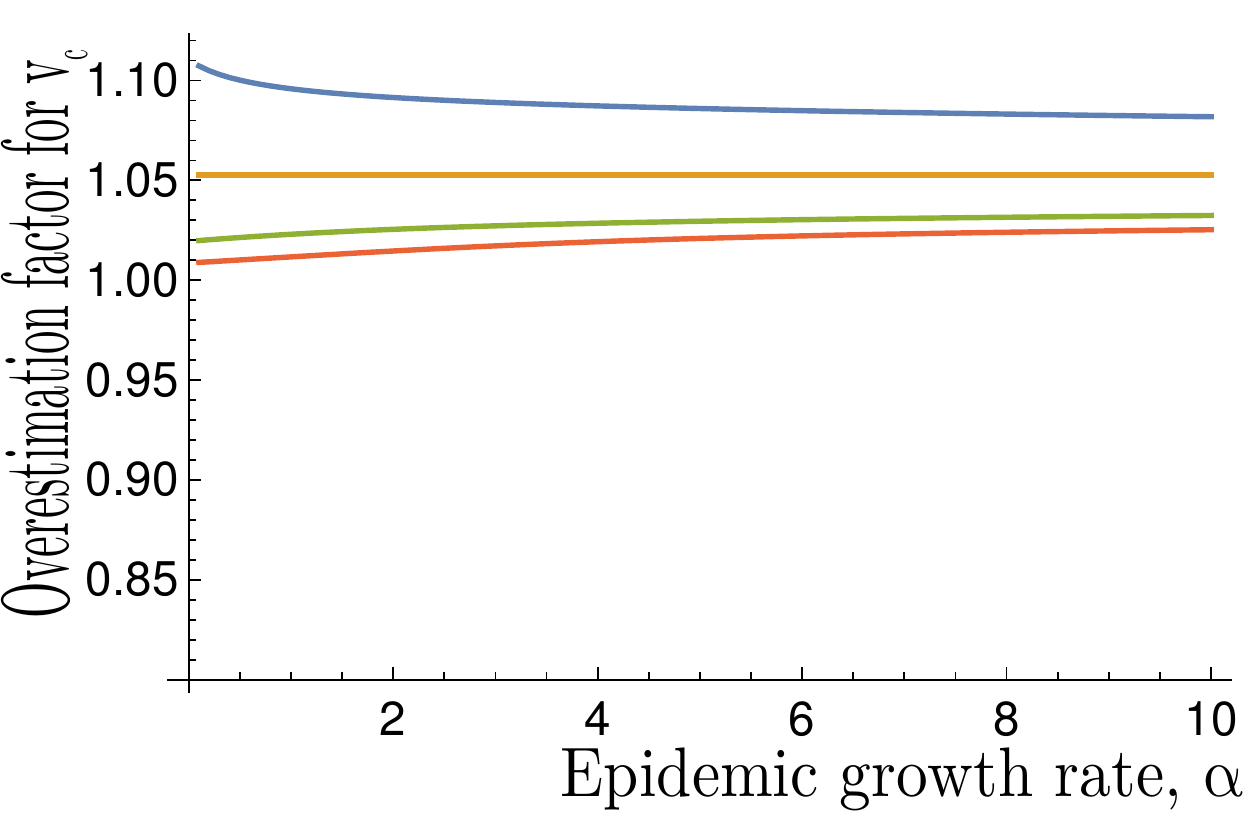}
}
\caption{The factor by which estimators based on homogeneous mixing will overestimate (a) the basic reproduction number $R_0$ and (b) the required control effort $v_c$ for the network case. Here the epidemic growth rate $\alpha$ is measured in multiples of the mean infectious period $1/\gamma$. The mean excess degree $\kappa= 20$. The infectious periods are assumed to follow a gamma distribution with mean 1 and standard deviation $\sigma\!=\!1.5$, $\sigma\!=\!1$, $\sigma\!=\!1/2$ and $\sigma\!=\!0$, as displayed from top to bottom. Note that the estimate of $R_0$ based on homogeneous mixing is $1\!+\!\alpha$. Furthermore, note that $\sigma\!=\!1$, corresponds to the special case of an exponentially distributed infectious period, while if $\sigma\!=\!0$, the duration of the infectious period is not random.}\label{ratiosfig}
\end{figure}

\subsection{Multi-type structure}
A second kind of infectious contact structure is multi-type structure. 
Often a community contains different types of individuals that display specific roles in contact behaviour. Types might be related to age-groups, social behaviour or occupation. It may be hard to classify all individuals into types and sometimes data on the types of individuals are missing. 
Furthermore, the number of parameters required to describe the contact rates between the types is large.  We assume that there are $K$ types of individuals, labelled $1,2, \cdots, K$, and that for $i=1, \cdots, K$ a fraction $\pi_i$ of the $n$ individuals in the population is of type $i$.
For the Markov SIR epidemic, we assume that the rate of close contacts from a given type-$i$ individual to a given type-$j$ individual is $\lambda_{ij}/n$. Note that here close contacts are not necessarily symmetric, i.e., if individual $x$ makes a close contact with individual $y$, then it is not necessarily the case that $y$ makes a close contact with $x$.  We assume again that individuals stay infected for an exponentially distributed time with expectation $1/\gamma$. The expected rate at which a given type-$i$ individual infects type-$j$ individuals at time $t$ since infection is $a_{ij}(t)=\lambda_{ij} \pi_j e^{-\gamma t}$. Here, $\lambda_{ij}/n$ is the rate at which the type-$i$ individual contacts a given type-$j$ individual, $n \pi_j$ is the number of type-$j$ individuals and $ e^{-\gamma t}$ is the probability that the type-$i$ individual is still infectious $t$ time units after being infected. It is well known  \cite{BC93,DHB12,Diek98,Done76} that the basic reproduction number $R_0= \rho_M$ is the largest eigenvalue of the matrix $M$, which has elements $m_{ij} = \int_0^{\infty} a_{ij}(t) dt,$ and the epidemic growth rate  $\alpha$ is such that $1=\int_0^{\infty} e^{-\alpha t} \rho_{A(t)} dt,$ where $\rho_{A(t)}$ is the largest eigenvalue of the matrix $A(t)$ with elements  $a_{ij}(t)$. Let $\rho$ be the largest eigenvalue of the matrix with elements $\lambda_{ij} \pi_j$ and note that $\rho_{A(t)} = \rho e^{-\gamma t}$. Therefore, 
\[1= \rho \int_0^{\infty} e^{-(\alpha+\gamma) t} dt \quad \mbox{and}   \quad R_0 =  \rho\int_0^{\infty} e^{-\gamma t} dt.\]
These equalities imply that
\[R_0 = \frac{\int_0^{\infty} e^{-\gamma t} dt}{\int_0^{\infty} e^{-(\alpha +\gamma) t}  dt} = 1 + \frac{\alpha}{\gamma},\]
which shows that the relation between $R_0$ and $\alpha$ for this class of multi-type Markov SIR epidemics is the same as for such an epidemic in a homogeneous mixing population (cf.\ equation (\ref{R0alpharel})). 

In Appendix A.5. we derive that estimators for $R_0$ and (if control measures are independent of the types of individuals) $v_c$  are \textit{exactly} the same as for homogeneous mixing in a broad class of SEIR epidemic models. This class includes the full epidemic model allowing for arbitrarily distributed latent and infectious periods and models in which the rates of contacts between different types keep the same proportion all of the time, although the rates themselves may vary over time  (cf.\ \cite{Diek98}).

\begin{figure}
\centering
\subfigure[ ]{
    \includegraphics[width=.5\textwidth]{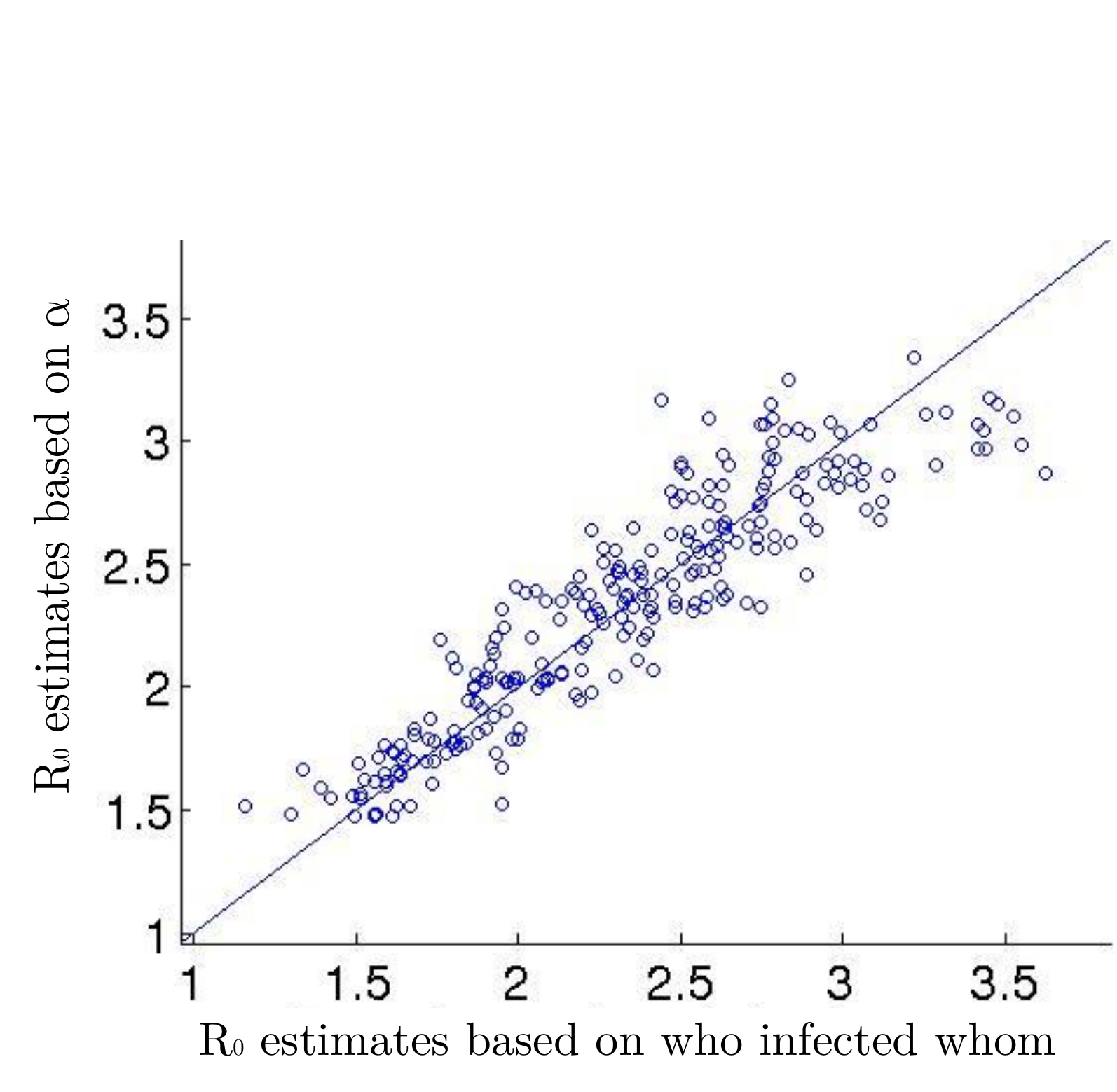}
}
\subfigure[ ]{
   \includegraphics[width=.2\textwidth]{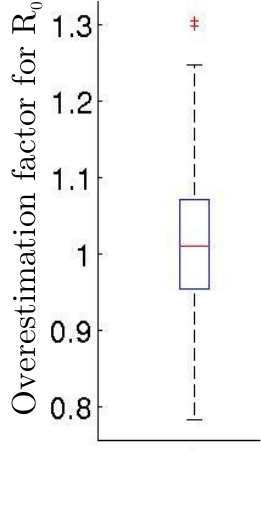}
}
\caption{The estimated basic reproduction number, $R_0$, for a Markov SEIR model in a multi-type population as described in \cite{Wall06}, based on the real infection process (who infected whom) plotted against the computed $R_0$, assuming homogeneous mixing, based on the estimated epidemic growth rate, $\alpha$, and given expected infectious period (5 days) and expected latent period (10 days). The infectivity is chosen at random, such that the theoretical $R_0$ is uniform between 1.5 and 3. The estimate of $\alpha$ is based on the times when individuals become infectious. In b) a box plot of the ratios is given.}\label{actvsestmult}
\end{figure}

We illustrate our findings on multi-type structures through simulations of SEIR epidemics in an age stratified population with known contact structure as described in \cite{Wall06}. Details on the population of approximately 14.6 million people, their types and contact intensities can be found in Appendix B. 
We use values of the average infectious period $1/\gamma$  and the average latent period $1/\delta$ close to the estimates for the 2014 Ebola epidemic in West Africa \cite{teamebola}. 
The simulation and estimation methods are described in detail in Appendix B.
We use two estimators for $R_0$. The first of these estimators is based on the average number of infections among the people who were infected early in the epidemic. This procedure leads to a very good estimate of $R_0$ if the spread of the disease is observed completely. 
The second estimator for $R_0$ is based on $\hat{\alpha}$, an estimate of the epidemic growth rate $\alpha$, 
and known expected infectious period $1/\gamma$ and expected latent period $1/\delta$, and is given by $(1+ \hat{\alpha}/\delta)(1+ \hat{\alpha}/\gamma)$. We calculate estimates of $R_0$ using these two estimators for 250 simulation runs. As predicted by the theory, the simulation results show that for each run the estimates are close to the actual value (Figure \ref{actvsestmult}(a)), without a systematic bias (Figure \ref{actvsestmult}(b)).

\subsection{Household structure}
\begin{figure*}[ht]
\centering
\subfigure[ ]{
   \includegraphics[width=.3\textwidth]{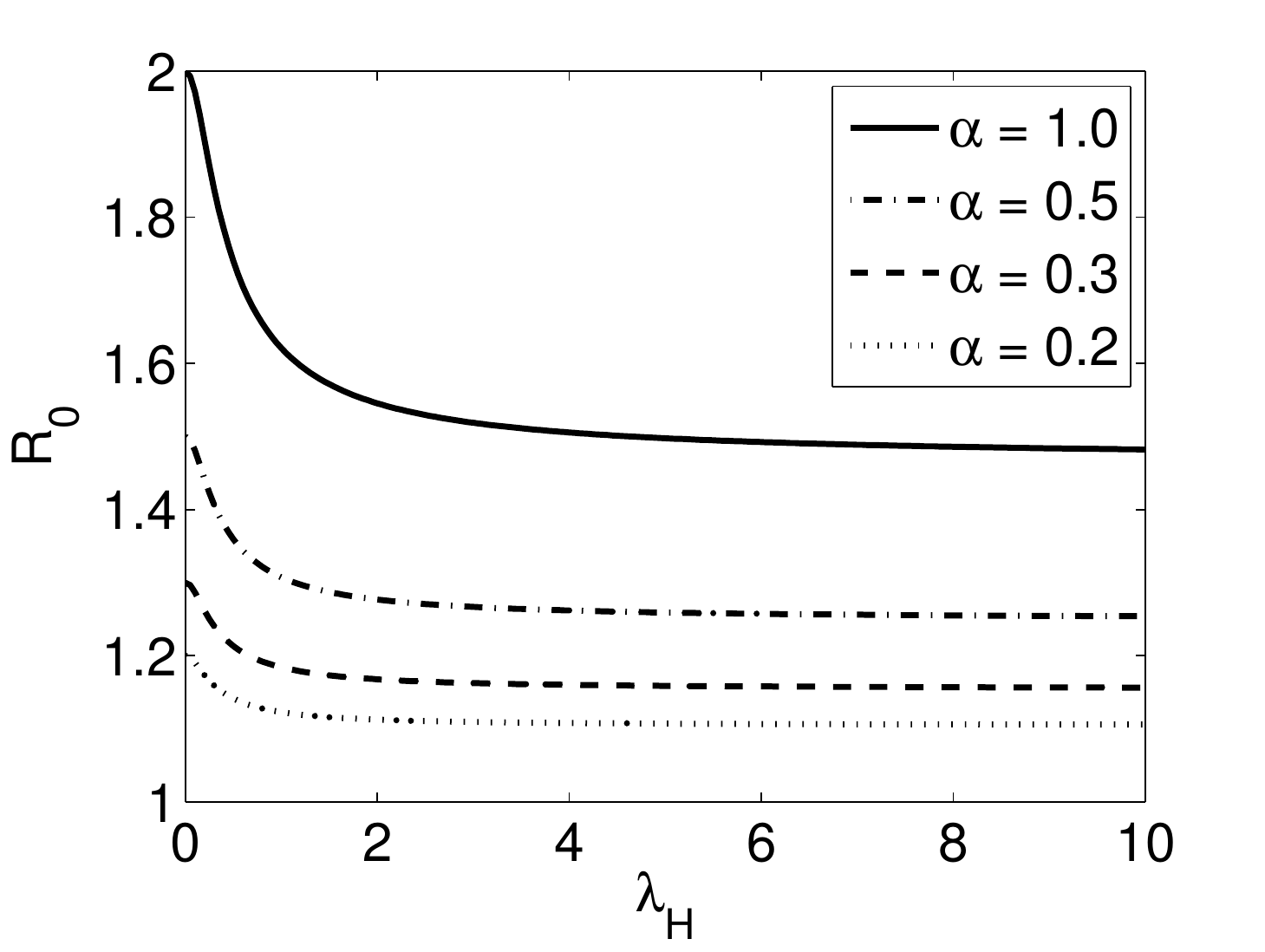}
}
\subfigure[ ]{
    \includegraphics[width=.3\textwidth]{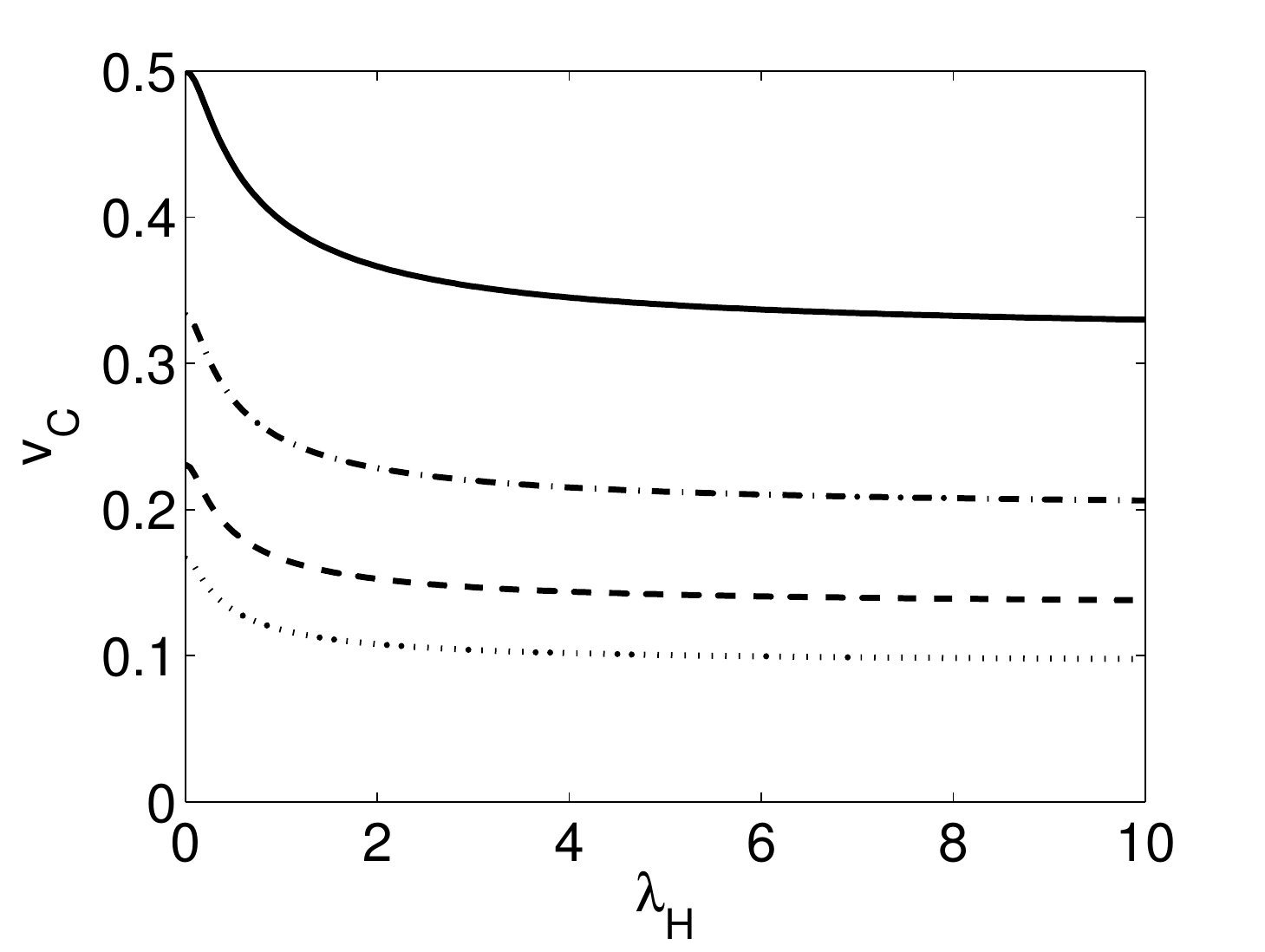}
}
\subfigure[ ]{
   \includegraphics[width=.3\textwidth]{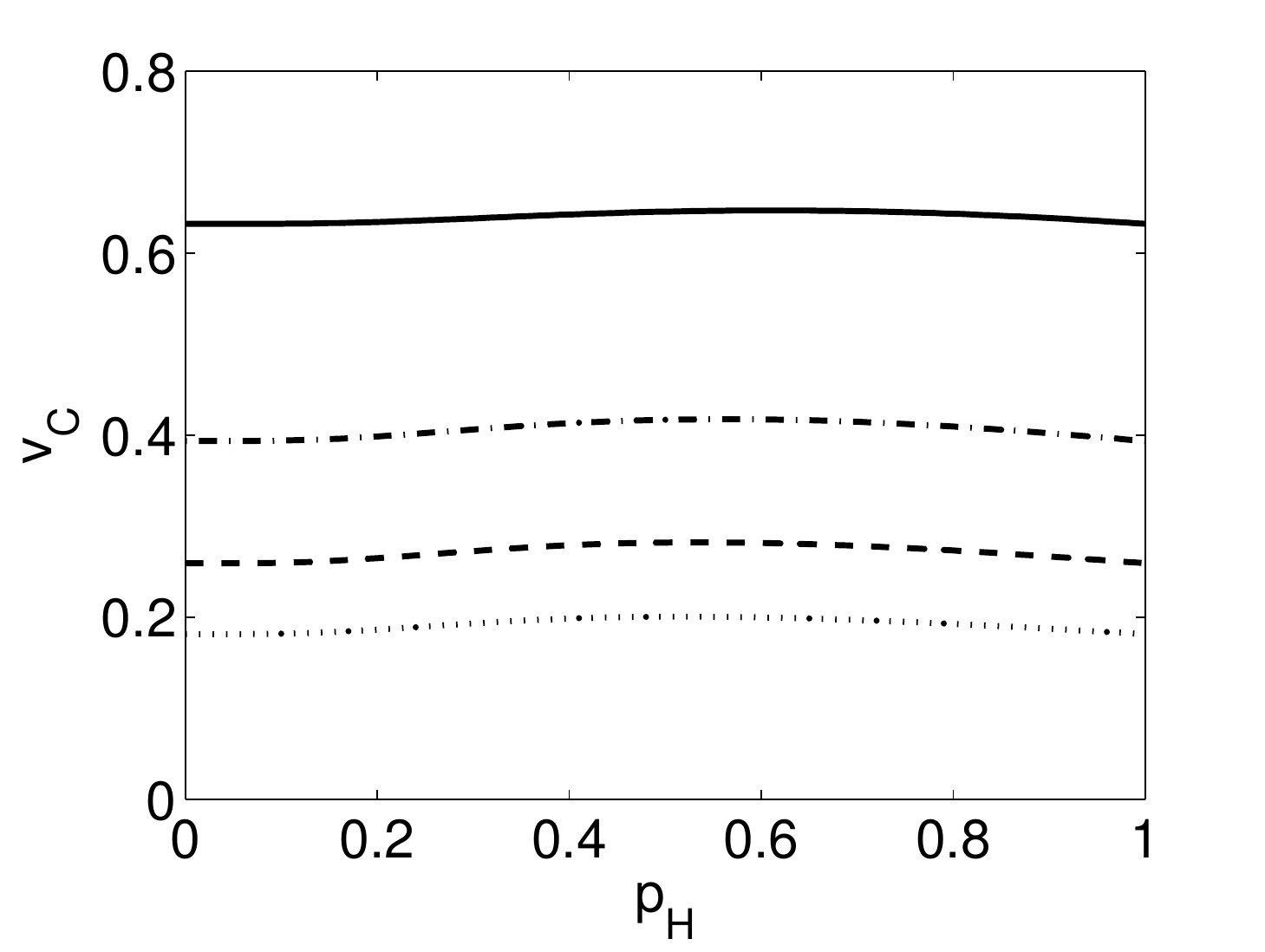}
}\\
\subfigure[ ]{
   \includegraphics[width=.3\textwidth]{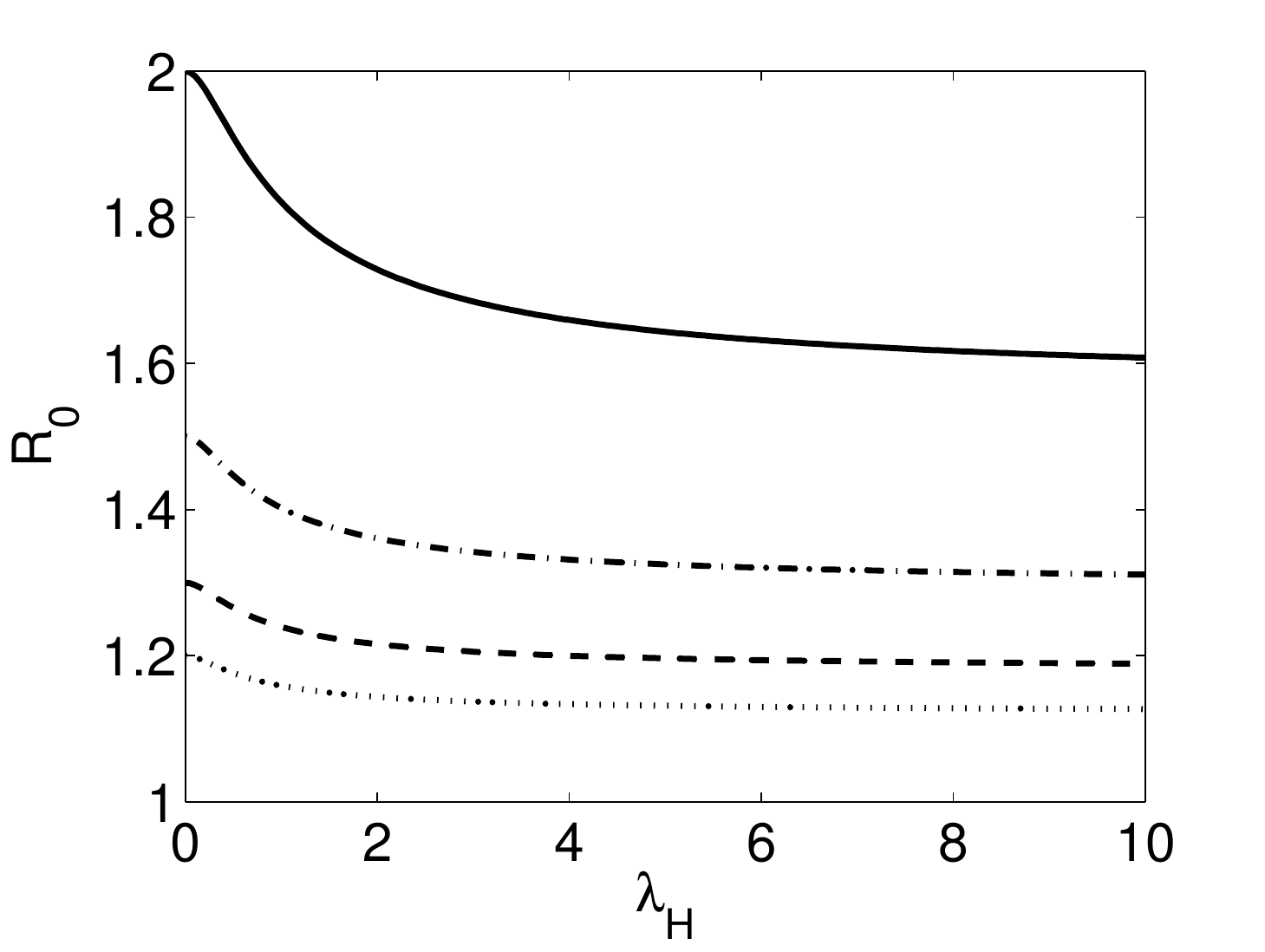}
}
\subfigure[ ]{
    \includegraphics[width=.3\textwidth]{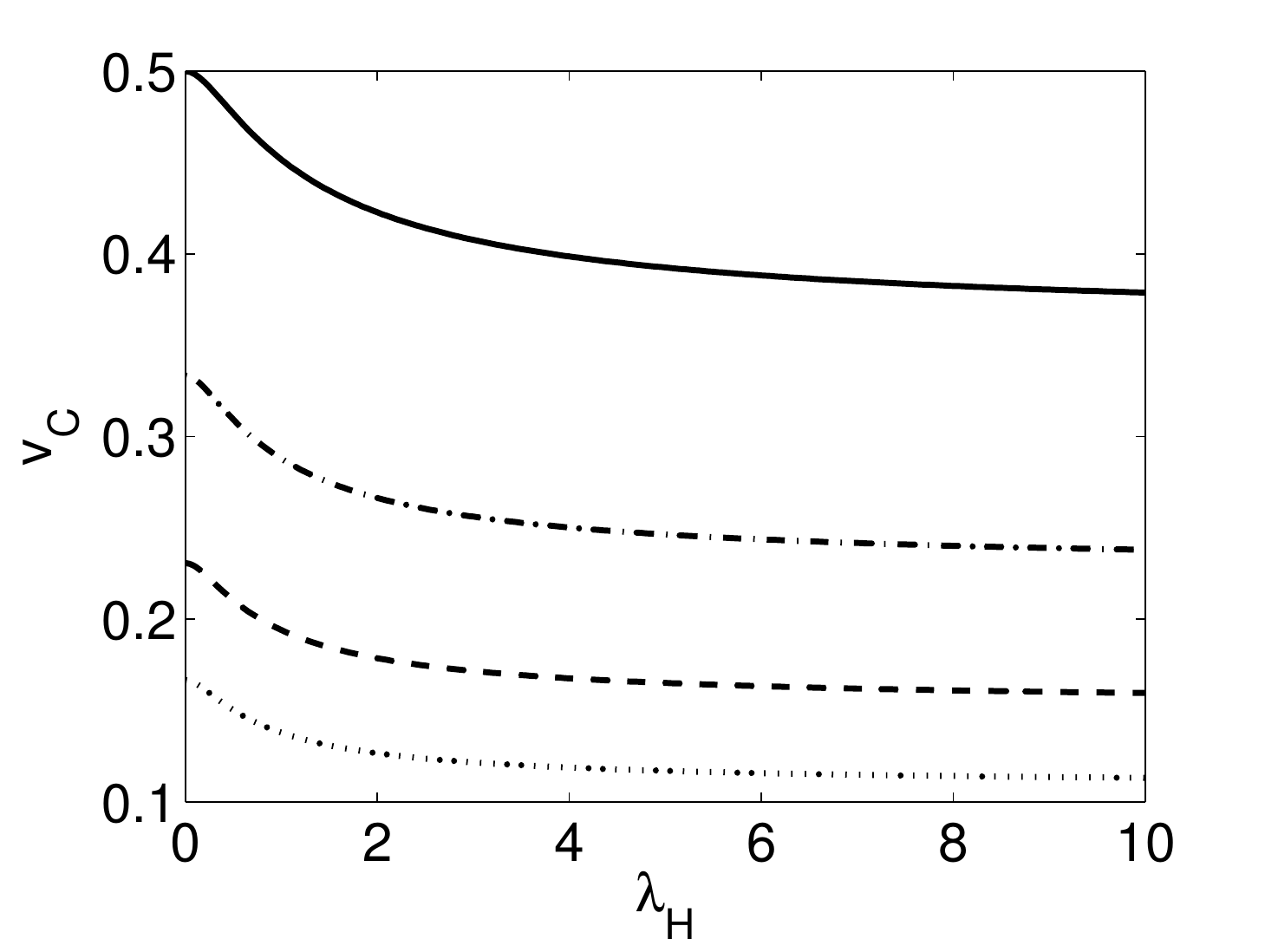}
}
\subfigure[ ]{
   \includegraphics[width=.3\textwidth]{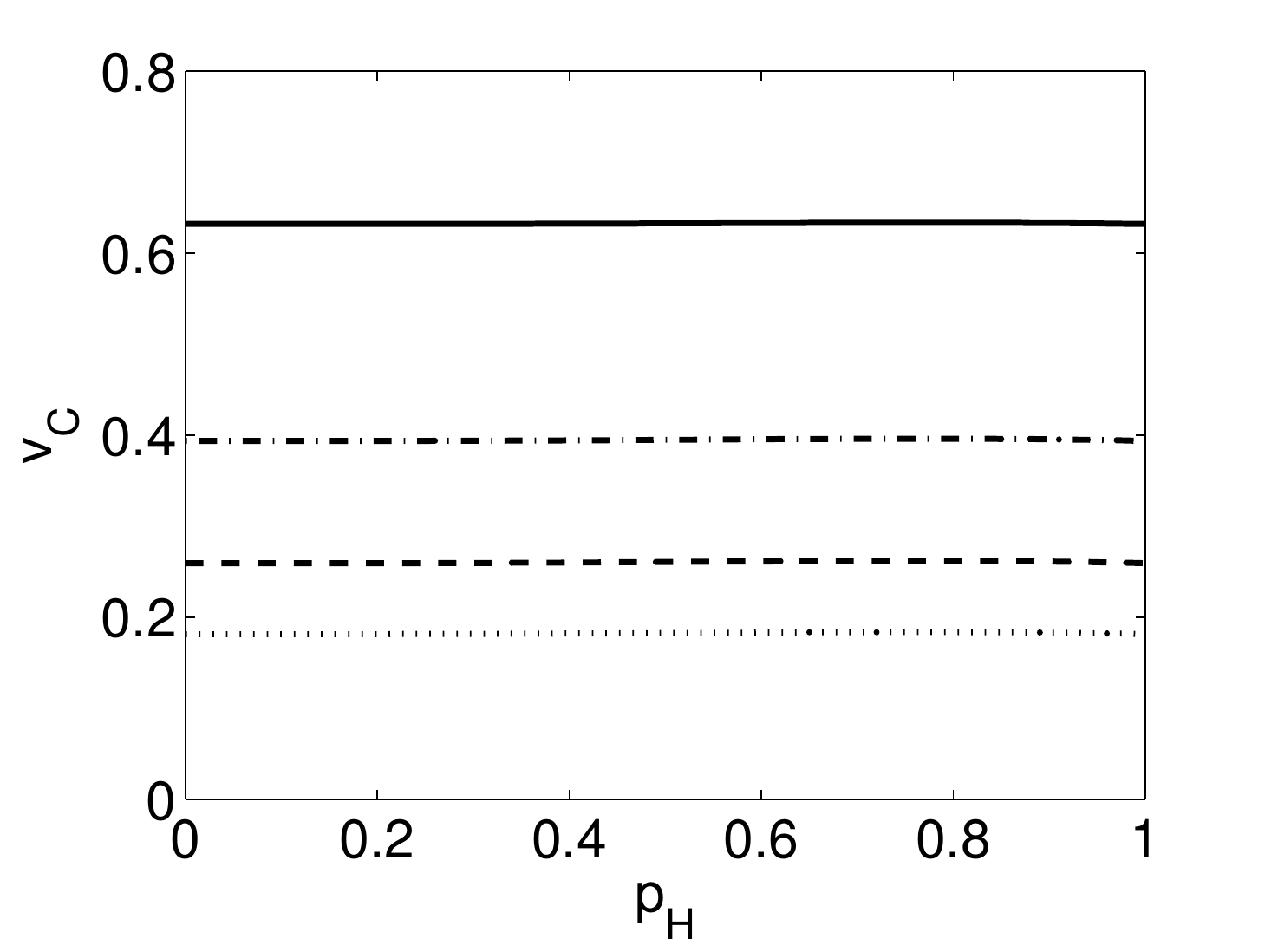}
}

\caption{Estimation of key epidemiological variables in a population structured by households. The basic reproduction number $R_0$ for Markov SIR epidemics with expected infectious period equal to 1 (a and d), critical vaccination coverage $v_c$  for Markov SIR epidemics (b and e)  and $v_c$ for Reed-Frost epidemics (c and f), as a function of the relative influence of within household transmission, in a population partitioned into households. 
For (a-c), 
the household size distribution is taken from a 2003 health survey in Nigeria \cite{nige2003} and is given by
$m_1 = 0.117, m_2=0.120, m_3=0.141, m_4=0.132, m_5=0.121, m_6= 0.108, m_7= 0.084, m_8=0.051, m_9= 0.126$; for  $i=1,2, \cdots, 9$, $m_i$ is the fraction of the households with size $i$.
For (d-f),
the Swedish household size distribution in 2013 taken from \cite{Statswed14}, is used and
is given by
$m_1 = 0.482, m_2=0.2640, m_3=0.102, m_4=0.109, m_5=0.01$.
The global infectivity is chosen so that the epidemic growth rate $\alpha$ is kept constant while the within household transmission varies. Homogeneous mixing corresponds to $\lambda_H=p_H=0$.}
\label{housepictures}
\end{figure*}

A third kind of infectious contact structure is household structure. This partitions a population into many relatively small social groups or households, which reflect actual households, school classes or workplaces. The contact rate between pairs of individuals from different households is small and the contact rate between pairs of individuals in the same household is much larger. This model was first analysed in detail in \cite{Ball97}. It is possible to define several different measures for the reproduction numbers for this model \cite{Beck95,G09}, but the best suited for our purpose is given in \cite{Pell12,Ball14}. For this model it is hard to find explicit expressions for $R_0$ and required control effort in terms of the observable epidemic growth rate. Numerical computations described in \cite{Ball14} suggest that the difference between the estimated $R_0$ based on $\alpha$ and the real $R_0$ might be considerable, but it is theoretically shown that the estimate is conservative for the most-commonly studied models. It is also argued that the required control effort $v_c \geq 1-1/R_0$ for this model, which implies that if we know $R_0$ and we base our control effort on this knowledge, we might fail to stop an outbreak. However, we usually do not have direct estimates for $R_0$ and even though it is not true in general that using $R_0$ leads to conservative estimates for $v_c$ \cite{Ball14}, numerical computations suggest that the approximation of $v_c$ using $\alpha$  and the homogeneous mixing assumption is often conservative.
This is shown in  Figure \ref{housepictures},
where we show estimates for $R_0$ and $v_c$ over a range of values for the relative contribution of the 
within-household spread. For each epidemic growth rate $\alpha$, the estimated values remain below the value obtained for homogeneous mixing (which corresponds to $\lambda_H=0$ and $p_H=0$, where $\lambda_H$ and $p_H$ are defined below).
We use two types of epidemics: in (a) and (b) the Markov SIR epidemic is used, while in (c)  the so-called Reed-Frost model is used, which can be interpreted as an epidemic in which infectious individuals have a long latent period of non-random length, after which they are infectious for a very short period of time. We note that for the Reed-Frost model the relationship between $\alpha$ and $R_0$ does not depend on the household structure (cf.\ \cite{Ball14}) 
and therefore, for this model, only the dependence of $v_c$ on the relative contribution of the within household spread is shown in Figure \ref{housepictures},
The household size distributions are taken from a 2003 health survey in Nigeria \cite{nige2003} and from  data on the Swedish household size distribution in \cite{Statswed14}. For Markov SIR epidemics, as the within-household infection rate $\lambda_H$ is varied, the global infection rate is varied in such a way that the computed epidemic growth rate $\alpha$ is kept fixed. For this model, $\alpha$ is calculated using the matrix method described in Section 4.1 of \cite{PFF11}.
For the Reed-Frost epidemic model, the probability that an infectious individual infects a given susceptible household member during its infectious period, $p_H$ is varied, while the corresponding probability for individuals in the general population varies with $p_H$ so that $\alpha$ is kept constant.
For this model, assuming that the unit of time is the length of the latent period, $R_0$ coincides with the initial geometric rate of growth of infection, so $\alpha= \log(R_0)$. From Figure \ref{housepictures}, we see that estimates of $v_c$ assuming homogeneous mixing are reliable for Reed-Frost type epidemics, although as opposed to all other analysed models and structures, the estimates are not conservative. We see also that for the Markov SIR epidemic, estimating $R_0$ and $v_c$ based on the homogeneously mixing assumption might lead to conservative estimates which are up to 40\% higher than the real $R_0$ and $v_c$.

The results obtained  for Markov SIR epidemics in the homogeneous mixing, network and multi-type structured population are summarized in Table \ref{tab:imp}. The results from household models are not in the table, since the expressions are hardly insightful.

\begin{table}[h]
 \caption{The epidemic growth rate $\alpha$, the basic reproduction number $R_0$ and required control effort $v_c$ for a Markov SIR epidemic model as function of model parameters in the homogeneous mixing, network and multi-type model and their relationship to each other.}
\begin{tabular}{lllll}
\hline
\ & Quantity of &  \multicolumn{2}{c}{Quantity of interest as function of}  & Ratio with \\
Model & interest & $\lambda$, $\gamma$ and $\kappa$ & $\alpha$, $\gamma$ and $\kappa$ & homogeneous mixing\\
\hline
homogeneous mixing & $\alpha$ & $\lambda-\gamma$ & - & - \\
\ & $R_0$ & $\frac{\lambda}{\gamma}$ & $1+\frac{\alpha}{\gamma}$ & -\\
\ & $v_c$ & $\frac{\lambda -\gamma}{\lambda}$ & $\frac{\alpha}{\alpha + \gamma}$ & -\\
\hline
network & $\alpha$ & $(\kappa -1) \lambda-\gamma$ & - & -\\
\ & $R_0$ & $\frac{\kappa \lambda}{\lambda + \gamma}$ &  $\frac{\gamma + \alpha}{\gamma + \alpha/\kappa}$ & 
 $1+\frac{ \alpha}{\gamma \kappa}$\\
\ & $v_c$ &  $1-\frac{\lambda +\gamma}{\kappa \lambda}$ & $\frac{\kappa -1}{\kappa}\frac{\alpha}{\alpha + \gamma}$ &  $1+ \frac{1}{\kappa-1}$\\
\hline
multi-type & $\alpha$ & $\gamma (\rho_{M}-1)$ &  - & -\\
\ & $R_0$ & $\rho_M$ &  $1+\frac{\alpha}{\gamma}$ & 1 \\
\ & $v_c$ &  $1-\frac{1}{\rho_M}$ & $\frac{\alpha}{\alpha + \gamma}$ & 1\\
\hline
\end{tabular}
\label{tab:imp}
\end{table}

\section{Estimation of $R_0$ and required control efforts for empirical network structure}

The three  kinds of infectious contact structure studied are caricatures of actual social structures. Those actual structures may contain features of all three caricatures, and reflect small social groups such as school classes and households in which individuals interact frequently, as well as distinct social roles such as those based on age and gender, and frequently repeated contacts among those acquaintances. This leads us to expect that estimators based on ignoring contact structure will in general result in a slight overestimation of $R_0$ and required control effort. 

\begin{figure}
\centering
\subfigure[ ]{
    \includegraphics[width=.5\textwidth]{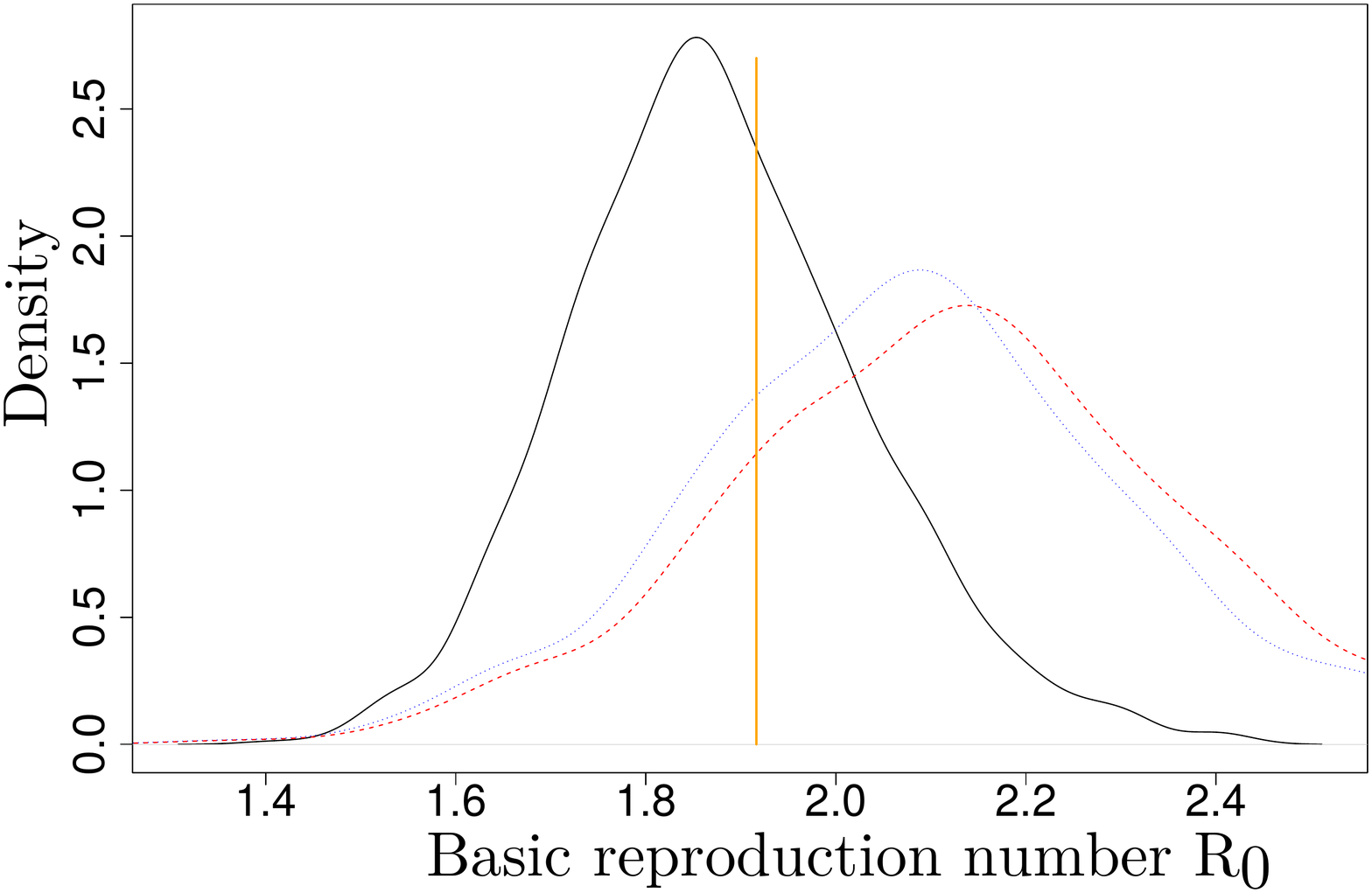}
}
\subfigure[ ]{
   \includegraphics[width=.2\textwidth]{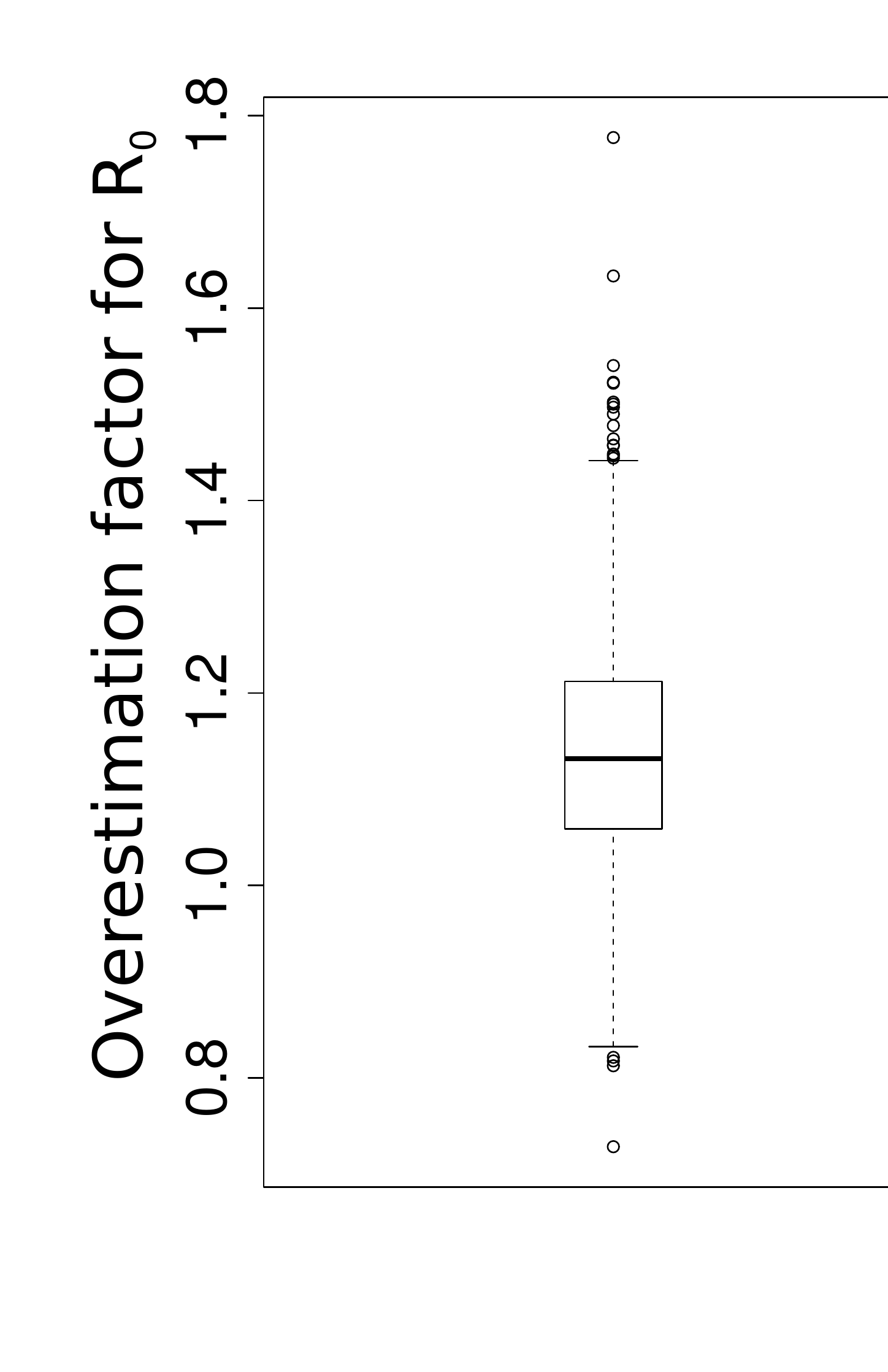}
}

\caption{Estimates for the basic reproduction number $R_0$ of an SEIR epidemic on the collaboration network in condense matter physics  \cite{Snap} based on 1000 simulated outbreaks. Each epidemic is started by 10 individuals chosen uniformly at random from the 23133 individuals in the population.
The infection rate is chosen such that $R_0 \approx 2$. In (a), the black line provides the density of estimates based on full observation of whom infected whom, the blue dashed line denotes the density of estimates based on the estimated epidemic growth rate $\alpha$ and the assumption that the network is a configuration model with known $\kappa$, while the red dotted line denotes the density of estimates based on $\alpha$ and the homogeneous mixing assumption.
The orange vertical line segment denotes the estimate of $R_0$ based only on the infection parameters and $\kappa$, assuming that the network is a configuration model (see equation (\ref{R0net}) in Appendix B).
We excluded the 50 simulations with highest estimated $\alpha$ and the 50 simulations with lowest estimated $\alpha$. In (b) a box plot of the ratios of the two $R_0$ estimates is provided.}
\label{Condmatpicture}
\end{figure}

We test this hypothesis further on some empirical networks taken from the Stanford Large Network Dataset collection \cite{Snap}.  In this report we present a network of collaborations in condense matter physics, where the individuals are authors of papers and authors are ``acquaintances'' if they were co-authors of a paper posted on the e-print service arXiv in the condense matter physics section between January 1993 and April 2004. In Appendix B we also analyse SEIR epidemics on two other networks from \cite{Snap}. The ``condense matter physics'' network is built up of many (overlapping) groups which represent papers. It was chosen since it is relatively large (23133 individuals and 93497 links), with over 92\% of the individuals in the largest component. The mean excess degree, $\kappa$, for this network is approximately 21 and small groups in which everybody is acquainted with everybody else are also present. 
In Figure \ref{Condmatpicture} we show the densities of estimates of $R_0$, based on 1000 simulations of an SEIR epidemic on this network, using parameters close to estimates for the spread of Ebola virus in West Africa \cite{teamebola}. The estimates are based on who infected whom in the real infection process (black line), the estimated epidemic growth rate and the configuration network assumption with $\kappa \approx 21$ (blue dashed line) and the estimated epidemic growth rate and the homogeneous mixing assumption (red dotted line). In most of the cases (886 out of 1000) the estimate of  $R_0$ based on homogeneous mixing is larger than the estimate based on who infected whom. In only 21 out of 1000 cases the estimate of $R_0$ based on homogeneous mixing is less than 90\% of the estimate of $R_0$ based on who infected whom. Half of the estimates of $R_0$ based on the epidemic growth rate and the homogeneous mixing assumption are between $12\%$ and $45\%$ larger than the estimate based on who infected whom. The difference in estimates might be explained through the relatively small average number of acquaintances per individual and the structure of small groups in which all individuals are acquaintances with all other individuals in the group.

\section{Discussion and conclusions}

In calculating the required control effort $v_c$, we have assumed that vaccinations, or other interventions against the spread of the emerging infection, are distributed uniformly at random in the population. For new, emerging infections this makes sense when we have little idea about the contact structure, and we do not know who is at high risk and who is at low risk of infection. When considering control measures that are targeted at specific subgroups, such as vaccination of the individuals at highest risk, closure of schools or travel restrictions, more information on infectious contact structure becomes essential to determine which intervention strategies are best. 
We note that for non-targeted control strategies the over-estimation of $R_0$ seems to be less for network structured and multi-type populations than for populations structured in households. Because, for epidemics among households better strategies than non-targeted control efforts are available \cite{Ball97, Ball02a,Beck97}, household (and workplace) structure is the first contact structure that should be taken into account.  

When the objective is to assess $R_0$ and $v_c$ from the observed epidemic growth rate of a new emerging infectious disease such as Ebola, ignoring contact structure leads to a positive bias in the estimated value. For both SIR epidemics and SEIR epidemics (Appendix A) this bias is small when the standard deviation of the infectious period is small enough compared to the mean as is the case for the Markov SEIR epidemic and even more so for the Reed-Frost model.
For Ebola in West Africa, we know that the standard deviation of the time between onset of symptoms, (which is a good indication of the start of the infectious period) and the time until hospitalization or death is of the same order as the mean. The same holds for the time between infection and onset of symptoms  \cite{teamebola}. These ratios of mean and standard deviation are well captured by the Markov SEIR epidemic. 

Our findings are important information for prioritizing data collection during an emerging epidemic, when assessing the control effort is a priority: it is most crucial to obtain accurate estimates for the epidemic growth rate from times of symptom onset of cases, and duration of the infectious and latent periods from data on who acquires infection from whom \cite{Lloy01,Nowa97,Robe07}. Data about the contact structure will be welcome to add precision, but  will have little effect on the estimated non-targeted required control effort in an emerging epidemic.

\pagebreak

\appendix
\section*{Appendices}

In these Appendices we discuss the mathematics behind some of the claims in the article, how simulations are performed and how estimates are obtained from the simulations. The parameters used for the Markov SEIR epidemic model are summarized in Table \ref{tab:parameters}.

\begin{table}[p]
 \centering
\begin{tabular}{ l  l}
 \hline
\multicolumn{2}{c}{general parameters and notation}\\
\hline
$\lambda$ & infection rate\\
$1/\gamma$ & average duration of infectious period\\
$1/\delta$ & average duration of latent period\\
$\alpha$ & exponential growth rate of number of infected individuals\\
$n$ &  population size\\
$R_0$ &  basic reproduction number, transmission potential,\\
\ & mean number of new infections caused by typical infected individual\\
$v_c$ &  required control effort, critical vaccination coverage\\
$I(t)$ & number of infectious individuals at time $t$\\
\hline
\multicolumn{2}{c}{parameters specific for network model}\\
\hline
$\mu$ & average number of acquaintances of individuals\\
$\sigma^2$ & variance of the number of acquaintances\\
$\kappa$ &  the mean number of acquaintances of newly infected individual,\\
\ & excluding the infector, $\kappa = \frac{\sigma^2}{\mu} + \mu-1$\\
\hline
\multicolumn{2}{c}{parameters specific for multi-type model}\\
\hline
$\iota$ & number of different types\\
$\pi_j$ & fraction of population with type $j$\\
$\lambda_{ij}$ & infection rate from type $i$ to type $j$ individual\\
$M$ & $\iota \times \iota$ next generation matrix, with elements $m_{ij} = \lambda_{ij}\pi_j/\gamma$\\
$J$ & $\iota \times \iota$ identity matrix\\
$\rho_A$ & largest eigenvalue of matrix $A$\\
\hline
\end{tabular}
\caption{Parameters and notation used for SEIR epidemic model in homogeneously mixing populations, on networks and in multi-type populations}
\label{tab:parameters}
\end{table}

\section{Mathematical methods}

\subsection{Introduction}

The stochastic and mathematical analysis of the spread of infectious diseases in large populations often relies on the theory of branching processes \cite{Jage75}. Branching processes are introduced as a model to describe family trees, where the simplifying assumption is that all women (in the branching process literature often the female lines are chosen) have the same probability, $p_k$, of having $k$ daughters, where $k$ can be any non-negative integer. Furthermore, the numbers of daughters of different  women are independent. 

It is clear that this model ignores important properties of real populations, such as changing circumstances which make the distribution of the number of children change over time and the fact that populations in general cannot grow indefinitely because of competition for resources. However, simple as it is, the model has proved useful in many situations. 

Branching processes are also useful to describe the spread of SEIR (susceptible $\to$ exposed $\to$ infectious $\to$ recovered/removed) epidemics, where an infection can be seen as a birth, with the infector being the mother and the infectee the daughter. 
In this model competition for resources is apparent, since once a susceptible individual is infected it cannot be infected again. However, if the population size $n$ is large and the number of no-longer-susceptible individuals is of smaller order than $\sqrt{n}$, then in homogeneous mixing populations, in configuration model network populations, in household models and in multi-type population models, suitable branching process approximations are very good (see e.g.\ \cite{BalDon1995}) 
and we use them without further justification. 
Branching processes can be analysed in real time and in generations. In real time, the Malthusian parameter or the  epidemic growth rate, $\alpha$ is arguably the most important parameter. A key theorem in branching processes \cite[Thm.6.8.1]{Jage75} states that if the number of women in the population grows large, then it roughly grows at a rate proportional to $e^{\alpha t}$, where $t$ is the time since the population began. From a generation perspective the essential parameter is $R_0$, which corresponds to the basic reproduction number or transmission potential in epidemic language.  This is the average number of daughters per typical woman (or number of infections per typical infectious individual in the epidemic setting). An outbreak can become large only if $R_0>1$, which happens if and only if $\alpha>0$. Note that if $R_0>1$, then it is still possible that the epidemic will go extinct quickly. The probability for this to happen can be computed \cite[Eq.\ 3.10]{DHB12} and is less than 1.

In the remainder of this appendix, we first discuss some useful results from the theory of branching processes. Then we apply them to epidemics in respectively homogeneously mixing populations, network populations, multi-type populations and household populations. Throughout we focus on $R_0$. It is however worth remarking that in homogeneously mixing populations, in (configuration model) network populations and in multi-type populations, we can deduce straightforwardly the required control effort or critical vaccination coverage, $v_c$ from $R_0$. For more extensive discussions on control effort and vaccination in the household model see \cite{Ball02a}.
We note that the critical vaccination coverage is based on vaccination uniformly at random, i.e.\ all people have the same probability of receiving the vaccine. As stated before, this vaccination strategy is not optimal if the population structure is known exactly, but since this relevant population structure is generally hard to obtain for emerging diseases, vaccination uniformly at random might be the best feasible method. 

Throughout we often use the superscripts ``(hom)'', ``(net)'', ``(mult)'', and ``(house)'', to refer to parameters and quantities associated with epidemics in respectively homogeneous mixing populations, network models, multi-type populations and populations consisting of households. 

As a leading example we use the Markov SEIR epidemic model. In this model pairs of individuals make (close) contacts independently at a rate which might depend on the pair (depending on the population structure). If an infectious individual contacts a susceptible one, the susceptible one becomes latently infected (exposed) and stays so for an exponentially distributed time with mean $1/\delta$, after which the individual becomes infectious. An individual stays infectious for an exponentially distributed time with mean $1/\gamma$, after which he or she is removed, which might mean that the individual dies, he or she recovers with permanent immunity or is isolated in a 100\% effective way. We also discuss the Markov SIR epidemic, in which there is no latent period (or $\delta = \infty$), but is the same as the Markov SEIR epidemic in all other respects.   We assume that there are only a few initially infective individuals in the population and all others are susceptible.

\subsection{Branching process results}

In this section we need some notation: for $t>0$, $\xi(t)$ is the random number of daughters a woman has given birth to by age $t$. Thus, $\xi(t)$  is a non-decreasing random process. Furthermore, define $\mu(t) = \mathbb{E}(\xi(t))$ as the expectation of $\xi(t)$. It is clear that $\mu(t)$ is also non-decreasing. For ease of exposition we assume that the derivative of $\mu(t)$ exists and is given by $\beta(t)$. Thus $\mu(t) = \int_{0}^t \beta(s) ds$. This assumption is not necessary and the results below can be generalized in a straightforward way to the case where $\mu(t)$ is not differentiable.  From the theory of branching processes \cite{Jage75}, we know that 
$R_0 = \mu(\infty) = \int_{0}^\infty \beta(s) ds$. 
In general there is no explicit expression for the Malthusian parameter $\alpha$, only the implicit equation specifying $\alpha$ 
\begin{equation}\label{maltdef}
1 = \int_{0}^{\infty} e^{-\alpha t} \beta(t) dt.
\end{equation}
If $R_0>1$ (the situation we are interested in), this equation has exactly one real positive solution \cite[p.\ 10]{Jage75}, and serves as a definition of $\alpha$. 

If the lifetime of a woman is distributed as the random variable $I$, and during her entire life she gives births to daughters at rate $\lambda$ (that is, the birth times of daughters form a homogeneous Poisson process with intensity $\lambda$), then $\beta(t) = \lambda \mathbb{P}(I>t)$. This gives that 
\begin{equation}\label{R0gen}
R_0 = \int_{0}^{\infty} \beta(t) dt =  \int_{0}^{\infty} \lambda \mathbb{P}(I>t) dt = 
\lambda \mathbb{E}(I).
\end{equation}
Here we have used the standard equality $\int_{0}^{\infty} \mathbb{P}(X>t) dt = \mathbb{E}(X)$ for any non-negative random variable $X$ (e.g. \cite[Sec.\ 4.3]{Grim92}).
From now on, for reasons of clarity, we assume that $I$ has a density which is denoted by $f_I(t)$. We may relax these assumptions without further consequences.
We deduce that 
\begin{multline}\label{alphagen}
1= \int_{0}^{\infty} e^{-\alpha t} \beta(t) dt=  
\int_{0}^{\infty} e^{-\alpha t} \lambda \mathbb{P}(I>t) dt= \lambda \int_{t=0}^{\infty}  \int_{s=t}^{\infty} e^{-\alpha t} f_I(s)ds dt\\ =
\lambda \int_{s=0}^{\infty}  \int_{t=0}^{s} e^{-\alpha t} f_I(s)dt ds = \frac{\lambda}{\alpha} \int_{0}^{\infty} (1-e^{-\alpha s}) f_I(s) ds\\ = 
\frac{\lambda}{\alpha} \mathbb{E}(1-e^{-\alpha I}) = \frac{\lambda}{\alpha} (1-\phi_I(\alpha)),
\end{multline}
where $\phi_I(\alpha)= \int_{0}^{\infty} e^{-\alpha t} f_I(t) dt = \mathbb{E}(e^{-\alpha I})$ is the Laplace transform of $I$ or, which is the same, the moment-generating function of $-I$. Equation (\ref{alphagen}) gives an implicit equation for $\alpha$.

If a woman only starts being fertile after a random ``latent'' period which is distributed as $L$ and has density $f_L(t)$, and after this period she is fertile for another, independent, period which is distributed as $I$, during which she gives birth to daughters at rate $\lambda$, then 
\[\beta(t) = \lambda \int_{0}^t f_L(u) \mathbb{P}(I>t-u) du ,\]
which is the convolution of $f_L(t)$ and $\beta_0(t)$, where $\beta_0(t)$ is the derivative of $\mathbb{E}(\xi(t))$ when the latent period is $0$. This  leads to  
\begin{multline}\label{R0gen2}
R_0 =  \int_{t=0}^{\infty} \lambda \int_{u=0}^t f_L(u) \mathbb{P}(I>t-u) du  dt\\ = \lambda \int_{u=0}^{\infty} \int_{t=u}^{\infty} f_L(u) \mathbb{P}(I>t-u) dt du  =
\lambda \mathbb{E}(I),
\end{multline}
where we have used the same computations as in (\ref{R0gen}). We note that $R_0$ is independent of the latent period.
Similarly we deduce that
\begin{multline}\label{alphagen2}
1 =  \int_{t=0}^{\infty} e^{-\alpha t} \lambda \int_{u=0}^t f_L(u) \mathbb{P}(I>t-u) du  dt \\ = \lambda \int_{u=0}^{\infty} \int_{t=u}^{\infty} e^{-\alpha t} f_L(u) \mathbb{P}(I>t-u) dt du \\= \lambda \int_{u=0}^{\infty} e^{-\alpha u} f_L(u) \int_{t=0}^{\infty} e^{-\alpha t} \mathbb{P}(I>t) dt du  = \frac{\lambda}{\alpha} (1-\phi_I(\alpha)) \phi_L(\alpha),
\end{multline}
where $\phi_L$ is the Laplace transform of the random variable $L$. If $L$ does not have a density the results above still hold.
Note that if $L=0$ with probability 1, then $\phi_L(\alpha) =1$ and we obtain (\ref{alphagen}) again.

\subsection{Homogeneously mixing populations}
\subsubsection{Constant infectivity}

For SEIR epidemics in a (homogeneously) randomly mixing population, every time an individual makes a close contact, it is with a random other individual from the population, which is chosen uniformly at random, independently of other close contacts. During the emerging phase of an epidemic it is unlikely that an individual is chosen, who is no longer susceptible. Thus, we assume that all close contacts of infectious individuals are with susceptible ones. To make the above mathematically fully rigorous, we should consider a sequence of epidemics in populations of increasing size and derive limit results for this sequence of epidemics \cite{BalDon1995}, but we leave out this level of technicality here.

If individuals each make close contacts independently at rate $\lambda^{(hom)}$, then we deduce from (\ref{R0gen2}) and (\ref{alphagen2}), that 
\[R^{(hom)}_0=\lambda^{(hom)} \mathbb{E}(I) \qquad \mbox{and} \qquad 1 = \frac{\lambda^{(hom)}}{\alpha} (1-\phi_I(\alpha)) \phi_L(\alpha).
\]

In particular,
\begin{equation}\label{homimp}
\frac{1}{R_0^{(hom)}}  =  \frac{(1-\phi_I(\alpha)) \phi_L(\alpha)}{\alpha \mathbb{E}(I)}.
\end{equation}
If $I$ is exponentially distributed with mean $1/\gamma$ and there is no latent period, then $\phi_I(\alpha) = \frac{\gamma}{\gamma + \alpha}$ and $\phi_L(\alpha)=1$, which leads to $R_0^{(hom)}= 1+ \alpha/\gamma$ as was deduced in (\ref{R0alpharel}). If the latent period is exponentially distributed with mean $1/\delta$, then $\phi_L(\alpha) = 
\frac{\delta}{\delta + \alpha}$. Thus in the Markov SEIR model, (\ref{homimp}) reads 
\[ \frac{1}{R_0^{(hom)}} = \frac{\gamma}{\gamma + \alpha} \frac{\delta}{\delta + \alpha}, 
\]
whence 
\[R_0^{(hom)}= \left(1+ \frac{\alpha}{\gamma}\right)\left(1+ \frac{\alpha}{\delta}\right)
\]

\subsubsection{Deterministic infectivity profile after latent period}

We proceed by considering the (non-Markov) SEIR model in which, during the infectious period $I$ being of random length, the close contact rate equals $h(\tau)$, where $\tau$ is the time since the infectious period starts. Note that we assume that $h(\tau)$ is non-random, i.e.\ identical for all infected individuals, but that the infectious period $I$ may end after a random time hence being different for different individuals. We also allow for a random latency period $L$ prior to the infectious period.
In this case, 
\begin{multline*}
R^{(hom)}_0 =  \int_{t=0}^{\infty} \int_{u=0}^t f_L(u) h(t-u) \mathbb{P}(I>t-u) du  dt\\
= \int_{u=0}^{\infty} \int_{t=u}^{\infty} f_L(u)  h(t-u) \mathbb{P}(I>t-u) dt du \\=  \int_{u=0}^{\infty}f_L(u) \int_{t=0}^{\infty}  h(t) \mathbb{P}(I>t) dt du =   \int_{0}^{\infty} h(t) \mathbb{P}(I>t) dt.
\end{multline*}
Similarly, we obtain
\begin{multline*}
1 =  \int_{t=0}^{\infty} e^{-\alpha t} \int_{u=0}^t f_L(u) h(t-u) \mathbb{P}(I>t-u) du  dt \\=  \int_{u=0}^{\infty} \int_{t=u}^{\infty} e^{-\alpha t} f_L(u)  h(t-u) \mathbb{P}(I>t-u) dt du \\= \int_{u=0}^{\infty}e^{-\alpha u} f_L(u) h(t) \int_{t=0}^{\infty} e^{-\alpha t}  \mathbb{P}(I>t) dt du  \\= \phi_L(\alpha) \int_{0}^{\infty} e^{-\alpha t} h(t) \mathbb{P}(I>t) dt,
\end{multline*}
whence,
\begin{equation}\label{homimpmore}
\frac{1}{R_0^{(hom)}} = \frac{\phi_L(\alpha) \int_{0}^{\infty} e^{-\alpha t} h(t) \mathbb{P}(I>t) dt}{\int_{0}^{\infty} h(t) \mathbb{P}(I>t) dt}.
\end{equation}
If $h(\tau) = \lambda$ is a  constant then this equality can be rewritten as (\ref{homimp}).

\subsection{Configuration model network populations}

\subsubsection{The network}

In this subsection we consider the configuration model network. In this network a fraction $d_k$ of the $n$ vertices (=individuals) has degree $k$, that is, a fraction $d_k$ of the population has $k$ other people it can have close contacts with, its acquaintances. The acquaintancies are represented by so-called bonds or edges. Out of all possible networks created in this way with given $n$ and $d_k$'s, we choose one uniformly at random. See \cite[Ch.3]{Durr06}, for more information on the construction of such networks.

We choose the (few) initial infective individuals all with equal probability (uniformly at random) from the population. If the population size $n$ is large, then the probability that an initially infective individual has $k$ acquaintances is $d_k$. However, by the construction of the network, the probability that an acquaintance of such an initially chosen infective has $k$ acquaintances is not $d_k$; for $k=1,2,\cdots$ the probability is given by
\[
\tilde{d}_k = \frac{k d_k}{\sum_{j=0}^{\infty} j d_j}=\frac{kd_k}{\mu},\ \text{ where } \mu=\sum_{j=0}^{\infty} jd_j,
\] 
since an initial infective is $k$ times as likely to be an acquaintance of an individual with degree $k$, than to be one of an individual with degree 1.
Now, if an individual is infected during the early stage of an epidemic, then at least one of its acquaintances is no longer susceptible (i.e.\ its infector). However, if $n$ is large, by the construction of the network the probability that its other acquaintances are still susceptible is close to 1. Hence, the expected number of susceptible acquaintances at the moment of infection of an individual infected during the early stages of the epidemic is 
\begin{equation}\label{defkappa}
\sum_{k=1}^{\infty} (k-1) \tilde{d}_k = \sum_{k=1}^{\infty} (k-1) \frac{k d_k}{\mu }= \frac{\sum_{k=0}^{\infty} (k-\mu)^2d_k}{\mu} + \mu - 1,
\end{equation}
which is equal to $\kappa$ as used in Section \ref{sec:netw}.

\subsubsection{The epidemic with constant infectivity}

Consider an SEIR epidemic on the configuration network described above. Assume again that $f_L(t)$ is the density of the duration of the latent period and $f_I(t)$ the density of the duration of the infectious period.
Assume that between every pair of acquaintances the rate of close contacts is $\lambda^{(net)}$ (i.e.\ close contacts occur according to independent Poisson processes with rate $\lambda^{(net)}$ per pair). The rate at which infection of a given acquaintance occurs at that time is $\lambda^{(net)}$ multiplied by the probability that the infector is infectious and has not previously infected this acquaintance, i.e. 
\[\lambda^{(net)} \int_{0}^t f_L(s) e^{-\lambda^{(net)} (t-s)} \mathbb{P}(I>t-s) ds.\]
 If the number of acquaintances of this infector is $k$, then the expected infectivity at time $t$ is 
\[(k-1) \lambda^{(net)} \int_{0}^t f_L(s) e^{-\lambda^{(net)} (t-s)} \mathbb{P}(I>t-s) ds.\]
Taking the mean over the number of acquaintances of an individual infected during the early stages of an epidemic, we obtain 
\[\beta(t) = \kappa \lambda^{(net)} \int_{0}^t f_L(s) e^{-\lambda^{(net)} (t-s)} \mathbb{P}(I>t-s) ds.\]
This leads, after manipulations as performed in (\ref{R0gen}) and (\ref{alphagen}), to
\begin{multline}
\label{firstarray}
R_0^{(net)} = \int_{0}^{\infty} \beta(t)dt = \int_{t=0}^{\infty}  \kappa \lambda^{(net)} \int_{u=0}^t f_L(u) e^{-\lambda^{(net)} (t-u)} \mathbb{P}(I>t-u) du dt \\  
=   \kappa \lambda^{(net)} \int_{u=0}^{\infty}  f_L(u)\int_{t=0}^{\infty} e^{-\lambda^{(net)} t} \mathbb{P}(I>t) dt du\\  
=   \kappa \lambda^{(net)} \int_{0}^{\infty} e^{-\lambda^{(net)} t} \mathbb{P}(I>t) dt  
= \kappa (1-\phi_I(\lambda^{(net)}))
\end{multline}
and
\begin{multline}
\label{secondarray}
1 = \int_{0}^{\infty} e^{-\alpha t} \beta(t)dt \\ = \int_{t=0}^{\infty}  e^{-\alpha t} \kappa \lambda^{(net)} \int_{u=0}^t f_L(u) e^{-\lambda^{(net)} (t-u)} \mathbb{P}(I>t-u) du dt \\ 
=   \kappa \lambda^{(net)} \int_{u=0}^{\infty} \int_{t=0}^{\infty} e^{-\alpha (t+u)} f_L(u) e^{-\lambda^{(net)} t} \mathbb{P}(I>t) dt du \\
=   \kappa \lambda^{(net)} \phi_L(\alpha) \int_{0}^{\infty}  e^{-(\alpha +\lambda^{(net)}) t} \mathbb{P}(I>t) dt\\   
= \kappa \phi_L(\alpha)  \frac{\lambda^{(net)}}{\alpha+\lambda^{(net)}} (1-\phi_I(\alpha+ \lambda^{(net)})).
\end{multline}

Combining these observations gives 
\begin{equation}\label{netimp}
\frac{1}{R_0^{(net)}} = \phi_L(\alpha) \frac{\lambda^{(net)}}{\alpha + \lambda^{(net)}}\frac{1-\phi_I(\alpha +\lambda^{(net)})}{1-\phi_I(\lambda^{(net)})}.
\end{equation}

If, as before, we consider the Markov SIR model in which $L=0$ and $I$ has an exponential distribution with mean $1/\gamma$, then (\ref{firstarray}) yields 
\begin{equation}
\label{R0net}
R_0^{(net)} = \kappa ((1-\phi_I(\lambda^{(net)}))= \kappa \frac{\lambda^{(net)}}{\lambda^{(net)}+\gamma}
\end{equation}
and (\ref{secondarray}) yields 
\[1 = \kappa  \frac{\lambda^{(net)}}{\alpha+\lambda^{(net)}} (1-\phi_I(\lambda^{(net)} + \alpha))= \kappa \frac{\lambda^{(net)}}{\lambda^{(net)} + \alpha+\gamma}.\]
The latter equality implies $\lambda^{(net)} = \frac{\gamma+\alpha}{\kappa-1}$, which inserted in the former gives \[R_0^{(net)} =  \frac{\gamma + \alpha}{\gamma + \alpha/\kappa}\]
as  claimed in Section \ref{sec:netw}.

If we consider the Markov SEIR epidemic in which the latent period has mean $1/\delta$ and the infectious period has mean $1/\gamma$, then $R_0^{(net)} = \kappa \frac{\lambda^{(net)}}{\lambda^{(net)}+\gamma}$ still holds, while (\ref{secondarray}) yields
\begin{equation}\label{netimp2}
1 = \frac{\delta}{\delta + \alpha}  \frac{\kappa \lambda^{(net)}}{\lambda^{(net)} + \alpha+\gamma},
\end{equation}
which in turn implies 
\[\lambda^{(net)} = \frac{(\gamma+\alpha)(\delta+\alpha)}{(\kappa-1)\delta - \alpha}.\]
Combining these observations gives that for the Markov SEIR epidemic
\[R_0^{(net)}= \frac{ \gamma+ \alpha}{ \gamma \delta/(\delta+\alpha)+ \alpha/\kappa }.
\]

\subsubsection{Deterministic infectivity profile after latent period}

As in the homogeneous mixing case we now assume that the infectivity, conditional upon still being infectious, is a function of the time $\tau$ since the infectious period starts, say $\hat{h}(\tau)$ (later we assume that $\hat{h}$ is proportional to $h$ as used in the homogeneous mixing population). Note that we assume that $\hat{h}(\tau)$ is not random, but that $L$ and $I$ are random and independent.
In this case, 
\begin{multline*}
R^{(net)}_0 =  \kappa \int_{t=0}^{\infty} \int_{u=0}^t f_L(u) \hat{h}(t-u) e^{- \int_{s=0}^{t-u} \hat{h}(s) ds} \mathbb{P}(I>t-u) du  dt \\ =  
\kappa \int_{u=0}^{\infty}  f_L(u)\int_{t=0}^{\infty} \hat{h}(t) e^{- \int_{s=0}^{t} \hat{h}(s) ds} \mathbb{P}(I>t) dt du \\
=  \kappa \int_{t=0}^{\infty} \hat{h}(t) e^{- \int_0^{t} \hat{h}(s) ds} \mathbb{P}(I>t) dt.
\end{multline*}
Similarly, we obtain

\begin{multline*}
1 = \kappa \int_{t=0}^{\infty} e^{-\alpha t} \int_{u=0}^t f_L(u) \hat{h}(t-u)  e^{- \int_{s=0}^{t-u} \hat{h}(s) ds} \mathbb{P}(I>t-u) du  dt \\
= \kappa \int_{u=0}^{\infty} e^{-\alpha u} f_L(u)\int_{t=0}^{\infty} e^{-\alpha t} \hat{h}(t) e^{- \int_{s=0}^{t} \hat{h}(s) ds} \mathbb{P}(I>t) dt du \\ = \kappa \phi_L(\alpha) \int_{t=0}^{\infty} \hat{h}(t)  e^{-\alpha t} e^{-\int_{s=0}^{t} \hat{h}(s) ds} \mathbb{P}(I>t) dt,
\end{multline*}
so,
\[
\frac{1}{R_0^{(net)}} = \phi_L(\alpha) \frac{\int_{t=0}^{\infty} \hat{h}(t) e^{-(\alpha t + \int_{s=0}^{t} \hat{h}(s) ds)} \mathbb{P}(I>t) dt}{ \int_{t=0}^{\infty} \hat{h}(t) e^{- \int_{s=0}^{t}\hat{h}(s) ds} \mathbb{P}(I>t) dt}.\]

\subsubsection{Comparison of $R_0^{(hom)}$ and $R_0^{(net)}$}

If we combine (\ref{homimp}) and (\ref{netimp}), and assume that $\alpha$ and the (constant) infection profiles (and thus $\phi_I$ and $\phi_L$) are known and the same for both models, then 
\begin{multline*}
\frac{R_0^{(hom)}}{R_0^{(net)}} = \frac{\frac{1}{(\alpha + \lambda^{(net)})\mathbb{E}(I)} (1-\phi_I(\alpha +\lambda^{(net)}))}{\frac{1}{\alpha \mathbb{E}(I)}(1-\phi_I(\alpha)) \frac{1}{\lambda^{(net)} \mathbb{E}(I)}(1-\phi_I(\lambda^{(net)}))}\\ = 
\frac{\mathbb{E}(I) \int_{0}^{\infty}  e^{-(\alpha + \lambda^{(net)})t } \mathbb{P}(I>t) dt}{\left(\int_{0}^{\infty} e^{-\alpha t} \mathbb{P}(I>t) dt\right)\left( \int_{0}^{\infty}  e^{- \lambda^{(net)}t } \mathbb{P}(I>t) dt\right)}.
\end{multline*}

To analyse this fraction, we introduce a random variable $Y$ by its distribution function
\[\mathbb{P}(Y \leq y) =\frac{ \int_{0}^{y} \mathbb{P}(I>t) dt}{\int_{0}^{\infty} \mathbb{P}(I>t) dt}, \quad \mbox{for $0 \leq y <\infty$.}\]
Using this and recalling that $\mathbb{E}(I)=\int_0^{\infty}\mathbb{P}(I>t)dt$, we can write
\[
\frac{R_0^{(hom)}}{R_0^{(net)}} = \frac{\mathbb{E}(e^{-\alpha Y} e^{-\lambda^{(net)}Y})}{\mathbb{E}(e^{-\alpha Y})\mathbb{E}(e^{-\lambda^{(net)}Y})}.
\]
Since $\lambda^{(net)}, \alpha>0$, we have that $e^{-\alpha x}$ and $e^{-\lambda^{(net)}x}$ are both non-increasing in $x$. Thus, by Chebyshev's integral inequality (or FKG inequality \cite[p.86]{Grim92}), we have that $e^{-\alpha Y}$ and $e^{-\lambda^{(net)}Y}$ are positively correlated, whence 
$R_0^{(hom)} \geq R_0^{(net)}$.

The difference between $R_0^{(hom)}$ and $R_0^{(net)}$ is small if $\kappa$ is relatively large compared to $R_0^{(hom)}$ and the standard deviation of the infectious period is not large compared to the mean. (See Figure 2). It can easily be seen that the opposite makes the approximation worse. Infections taking place a long time after the start of an infector's infectious period  contribute relatively little to $\alpha$; on the other hand all infections make the same contribution to $R_0$.
Also note, that if in the network model a given individual infects all of his/her acquaintances with large probability (say 99\%) if he/she is infectious for a middle-long time (say $T$), then increasing the infectious period to $2T$ has little effect on the epidemic both on its size (which relates to $R_0$) and its speed (which relates to $\alpha$). However, in a homogeneously mixing model, the offspring (which contributes to $R_0$) would double in expectation in this situation, while the speed of the epidemic would hardly change.
Thus, if the standard deviation of the infectious period is large, we cannot ignore the large infectious periods which cause the discrepancy between $R_0^{(hom)}$ and $R_0^{(net)}$.

Now consider the second special case discussed above: the infectivity profile, conditional upon still being infectious, $\hat{h}(\tau)$ is not constant, but is proportional to $h(\tau)$ for the homogeneous mixing model, where $\tau$ is the time since an individual starts to be infectious. Let $\lambda := \hat{h}(\tau)/h(\tau)$. Then,
\[
\frac{R_0^{(hom)}}{R_0^{(net)}} = \frac{\int_{0}^{\infty} h(t) \mathbb{P}(I>t) dt}{\int_{0}^{\infty} e^{-\alpha t} h(t) \mathbb{P}(I>t) dt}\frac{\int_{t=0}^{\infty} \lambda h(t) e^{-(\alpha t + \lambda \int_{\tau=0}^{t} h(\tau ) d\tau)} \mathbb{P}(I>t) dt}{ \int_{t=0}^{\infty} \lambda h(t) e^{- \lambda \int_{\tau=0}^{t} h(\tau ) d\tau } \mathbb{P}(I>t) dt}.\]

As for the SEIR model with constant rates, we introduce a random variable $Y'$ by its distribution function
\[\mathbb{P}(Y' \leq y) =\frac{ \int_{0}^{y} h(t) \mathbb{P}(I>t) dt}{\int_{0}^{\infty} h(t) \mathbb{P}(I>t) dt}, \quad \mbox{for $0 \leq y <\infty$.}\]
Using this we can write
\begin{equation}
\frac{R_0^{(hom)}}{R_0^{(net)}} = \frac{\mathbb{E}(e^{-\alpha Y'} e^{-\lambda \int_{0}^{Y'} h(\tau ) d\tau })}{\mathbb{E}(e^{-\alpha Y'})\mathbb{E}(e^{-\lambda \int_{0}^{Y'} h(\tau ) d\tau})}. \label{ratioR0cons}
\end{equation}
Since $\lambda$ and $\alpha$ are positive and $h(\tau)$ is a non-negative function, we have that $e^{-\alpha x}$ and $e^{-\lambda \int_{\tau=0}^x h(\tau ) d\tau}$ are both non-increasing in $x$. Thus, copying the argument above, we have that 
$R_0^{(hom)} \geq R_0^{(net)}$.
We note that although (\ref{ratioR0cons}) does not explicitly depend on $\kappa$, the relationship between $\alpha$ and $\lambda$ and $h(\tau)$ does and therefore the exact value of the right hand side does as well.

\subsubsection{Example of a model where $R_0^{(hom)} < R_0^{(net)}$}

The result $R_0^{(hom)}\geq R_0^{(net)}$ does not hold in general if $h(\tau )$ is a random function instead of a deterministic function, i.e.\ $h(\tau )$ is different for different people, following some distribution over stochastic processes.
This is shown in the following extreme example. 

We assume that every infective individual is infectious for exactly one point in time, at which he/she infects a random number of other individuals.
In the homogeneous mixing case, with probability $1/3$ an infectious individual infects on average 2 other individuals at time 0 (relative to his/her time of infection), while with probability $2/3$ he/she infects on average 1 other individual at time 1. 
This corresponds to \[\mu(t) =  2\frac{1}{3} + \frac{2}{3}\ind(t \geq 1),\]
leading to $R_0^{(hom)}=4/3$ and $1= 2/3 + (2/3) e^{-\alpha}$, which implies $e^{-\alpha} = 1/2$ (or $\alpha = \log[2]$).

In the corresponding network case we assume every individual has 3 acquaintances, so $\kappa=2$. With probability $1/3$ an infectious individual infects each of his/her susceptible acquaintances with probability $1-e^{-2\lambda}$ independently at time 0, while with probability $2/3$ he/she 
infects each of his/her susceptible acquaintances with probability $1-e^{-\lambda}$ independently at time 1. Here $\lambda$ is chosen such that $e^{-\alpha} = 1/2$.

For this model $\mu(t) = 2 \left[(\frac{1}{3}(1-e^{-2\lambda}) + \frac{2}{3}(1-e^{-\lambda}) \ind(t>1)\right],$
leading to the equations
\[R_0^{(net)}= \frac{2}{3}(1-e^{-2\lambda}) + \frac{4}{3}(1-e^{-\lambda})\qquad \mbox{and} \qquad 1= \frac{2}{3} ((1-e^{-2\lambda}) + (1-e^{-\lambda})).\]
Some algebra gives that $e^{-\lambda} = \frac{\sqrt{3}-1}{2}$, which implies 
\[R_0^{(net)}= 2-\frac{\sqrt{3}}{3} >\frac{4}{3} = R_0^{(hom)}.\]

\subsection{Multi-type epidemics}\label{multsec}

For the SEIR epidemic in a multi-type population, we assume that there are $\iota$ types of individuals, labelled $1,2, \cdots, \iota$ and again that the population is large. Additionally we assume that the number of individuals of each type is large, and in what follows we assume that there is no relevant depletion of susceptibles of any type during the initial stages of the epidemic. We assume that a fraction $\pi_i$ of the community is of type $i$. 
Furthermore, we assume that not all close contacts lead to infection. However, we do assume that the probability that a close contact between a susceptible and an infectious individual leads to infection depends only on the time since infection of the infectious one, $\tau$. This probability is random (i.e.\ different for different individuals) and is denoted by $\Lambda(\tau)$. Note that we assume that the distribution of $\Lambda(\tau)$ does not depend on the types of the individuals. The random function $\Lambda$ incorporates the latent and recovered period, in the sense that before the end of the latent period and after recovery $\Lambda(\tau) =0$. We use $g(\tau) = \mathbb{E}(\Lambda(\tau))$ for the expected probability of infection at age $\tau$ of a randomly selected individual. In an SIR epidemic the infectivity is often a function of $\tau$ conditioned on the individual still being infectious at time $\tau$. In that case $g(\tau)$ can be written as $h(\tau)\mathbb{P}(I>\tau)$.
Close contacts are not necessarily symmetric. That is,  if individual $x$ makes a close contact with individual $y$, then it is not necessarily the case that $y$ makes a close contact with $x$. The rate of close contacts from a given type $i$ individual to a given type $j$ individual is $\lambda_{ij}/n$. Therefore the expected number of $j$-individuals that an infected $i$-individual infects up to its ``age'' (time since infection)  $t$ during the early stages of an outbreak when all individuals are susceptible is given by
\begin{equation}
m_{ij}(t)= \int_{0}^{t} a_{ij}(\tau) d\tau, \qquad \mbox{where} \qquad a_{ij}(\tau)= \lambda_{ij}\pi_j g(\tau).
\end{equation}
The matrices $M(t)$ and $A(t)$ are defined by respectively $M(t) = ( m_{ij}(t) ) $ and $A(\tau) = (a_{ij}(\tau))$. Furthermore, we define $M=M(\infty) = \left(m_{ij}(\infty)\right)$ as the next generation matrix.
It is well-known that the basic reproduction number $R_0^{(mult)}$ is given by the dominant (i.e.\ ``largest'') eigenvalue of $M$, also denoted by $\rho_M$ \cite{DHB12,Diek98}.

To determine the epidemic growth rate, $\alpha$, we use Equation (6.4) and the subsequent paragraphs from \cite{Diek98}. This translates into that the dominant eigenvalue of $\int_0^{\infty} e^{-\alpha \tau} A(\tau) d\tau$ should equal 1, where the integral is taken elementwise. Now we use that  
\begin{multline*}
\int_{0}^{\infty} e^{-\alpha \tau} a_{ij}(\tau) d\tau = \int_{0}^{\infty} e^{-\alpha \tau} \lambda_{ij}\pi_j g(\tau) d\tau  = \lambda_{ij}\pi_j \int_{0}^{\infty} e^{-\alpha \tau}  g(\tau) d\tau\\= \frac{\int_{0}^{\infty} e^{-\alpha \tau}  g(\tau) d\tau}{\int_{0}^{\infty}  g(\tau) d\tau}m_{ij}(\infty).\end{multline*}
Hence, $\rho_A$, the largest eigenvalue of the matrix  $\int_{0}^{\infty} e^{-\alpha \tau} A(\tau) d\tau$ is given by $\rho_M$ multiplied by  $\int_{0}^{\infty} e^{-\alpha \tau}  g(\tau) d\tau/(\int_{0}^{\infty}g(\tau) d\tau)$, where $\rho_M$ is the largest eigenvalue of $M$.
In particular this gives that \[
\frac{1}{R_0^{(mult)}} = \frac{\int_{0}^{\infty} e^{-\alpha \tau}  g(\tau) d\tau}{\int_{0}^{\infty}g(\tau) d\tau}.\]

Notice that in the homogeneous case, i.e. the case with $\iota=1$ and  \[\mu(dt) = g(t) \lambda_{11} dt,\] we get the same relationship between  $\alpha$ and $R^{(hom)}_0$ (as given in equation (\ref{homimpmore}), with $h(\tau) \mathbb{P}(I>\tau)=g(\tau) \lambda_{11}$)
as between $\alpha$ and $R^{(mult)}_0$, which implies that ignoring the population structure does not affect the estimates for $R_0$.

\subsection{Household epidemics}

Household epidemics are harder to study in this context (compared to homogeneous, network and multi-type epidemics) and already several papers are dedicated to these epidemics, e.g.\ \cite{Ball97}.
In particular, there is no easy way to compute $R_0$ or $\alpha$ (instead other threshold parameters are often derived). Furthermore, if $v_c$ is the critical vaccination coverage when vaccination is applied uniformly at random (i.e.\ the required control effort), then the relationship \[v_c^{(house)} = 1-1/R_0^{(house)}\] does not hold in general. Also, if the household structure is observed, then there are better vaccination strategies than vaccination uniformly at random \cite{Ball02a}. (The same is true if the degrees of individuals are observed in the network model and if the types of individuals and their relative infectivities and susceptibilities are known in the multi-type model). However, in this article we consider the case where the population structure is hard to obtain. In that case vaccination uniformly at random seems to be the most natural vaccination strategy. Reproduction numbers for household epidemics and the relationships with vaccination uniformly at random and the epidemic growth rate are studied in great detail in \cite{Ball14} and some of the results will be repeated here.

For the household model we assume that the population is partitioned in $n/m$ households (or groups or cliques) of equal size $m$. So, we assume that $n$ is an integer multiple of the positive integer $m$. For a population where the households are not of equal size we refer 
to \cite{Pell12}. We consider only SEIR models in which individuals have constant infectivity during their infectious period. Individuals contact each other with global contacts at per-pair rate $\lambda_G/n$, while members of the same household make additionally local contacts at per-pair rate $\lambda_H$. Note that, unlike in Section \ref{multsec}, we assume that close contact of an infective with a susceptible necessarily results in the infection of the latter. 

We use the basic reproduction number $R_0^{(house)}$ as defined in \cite{Pell12,Ball14}, since this is the parameter having interpretation closest to the common $R_0$ definition. This $R_0^{(house)}$ can be computed by considering one isolated household of size $m$, which has one initial infectious individual and $m-1$ susceptibles. Let $\mu_0 =1$ and let  $\mu_1$ be the expected number of individuals in this household with whom the initial infective makes close
contact during its infectious period (the first generation). Similarly $\mu_i$ is the expected number of individuals in the $i$-th generation, that is, the expected number of initially susceptible individuals which were not in the first $i\!-\!1$ generations, but have a close contact with a generation $(i\!-\!1)$ individual during its infectious period. Note that $\mu_i =0$ for $i \geq n$.
In \cite{Pell12} it is shown that $R_0^{(house)}$ is the unique positive $x$ which solves \[1= \lambda_G \mathbb{E}(I) \sum_{i=0}^{m-1} \frac{\mu_i}{x^{i+1}}.\] If the households are not all of the same size then the $\mu_i$ are replaced by household-size-biased averages, see Section 3.3.\ of \cite{Pell12}.

In Section 2.6 of \cite{Ball14} it is shown that for SEIR epidemics $R_0$ estimates based on $\alpha$ and the homogeneous mixing assumption are conservative. We note that $\alpha$ is in general implicitly defined as the solution of an equation involving the infectivity profile of a household. Further 
arguments provided in \cite{Ball14} also show that in general \[v_c^{(house)}\!\geq\!1\!-\!1/R_0^{(house)}.\] 
If we estimate $v_c$ based on $\alpha$ and the homogeneous mixing assumption, then in most numerically analysed cases enough people are vaccinated. However, some counter examples are provided in \cite{Ball14}.

In Figure 4 the dependence of $R_0$ and $v_c$ on the relative contribution of the within household spread is illustrated for a household size distributions taken from Nigerian and Swedish datasets \cite{nige2003,Statswed14}.

\section{Simulations}

\begin{figure}[ht]

\centering
\subfigure[ ]{
    \includegraphics[width=.48\textwidth]{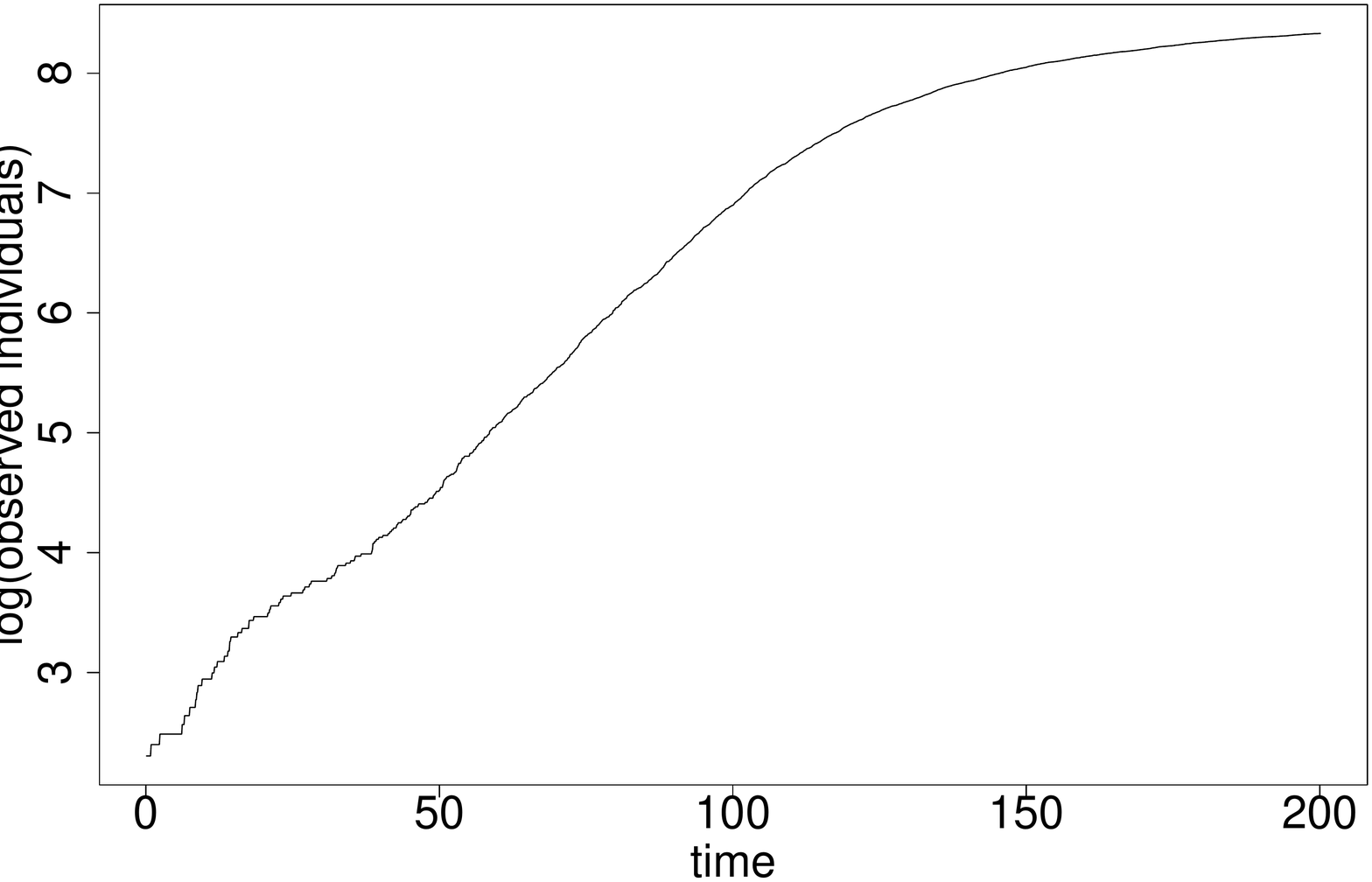}
}
\subfigure[ ]{
   \includegraphics[width=.48\textwidth]{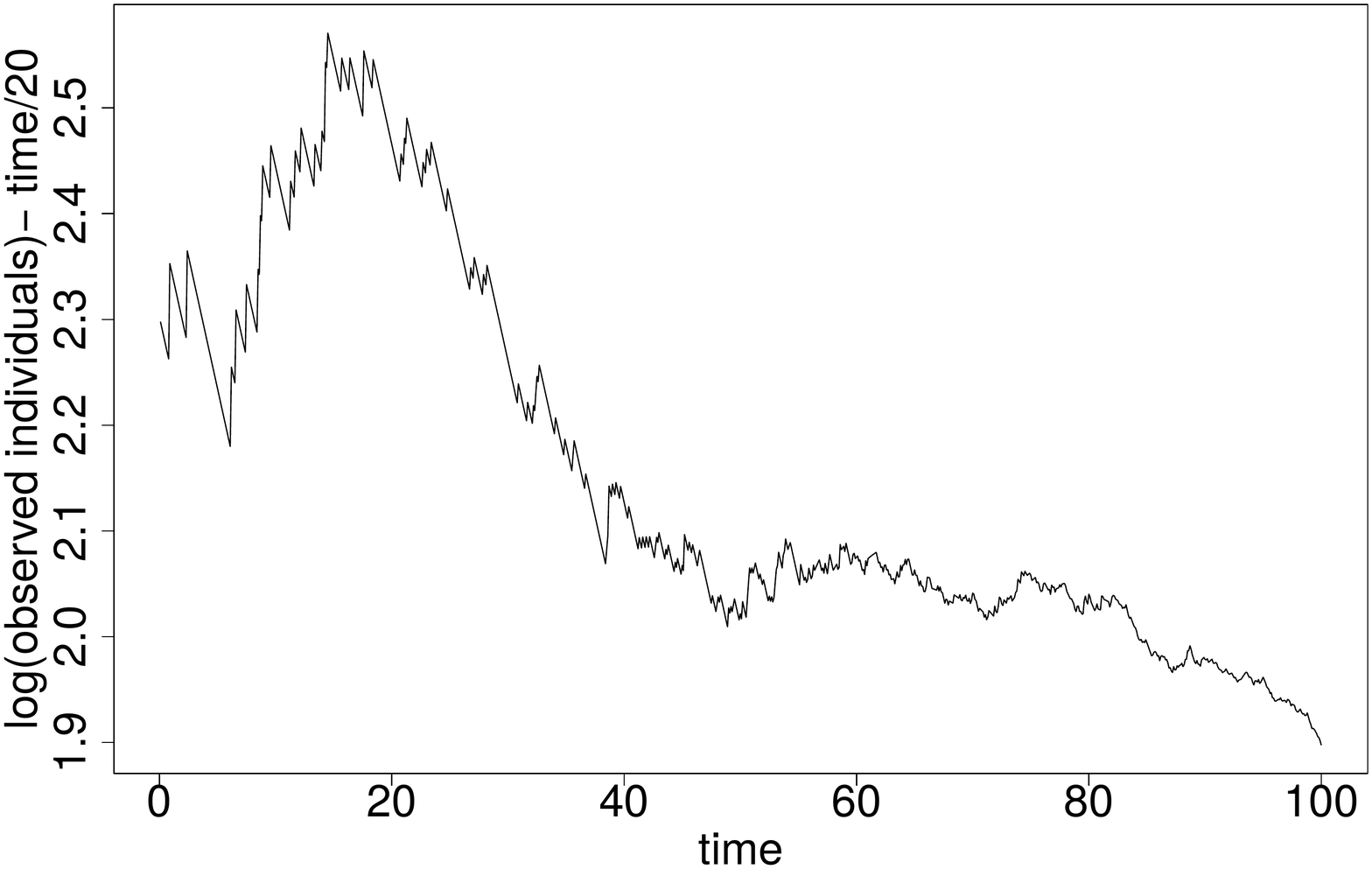}
}

\caption{(a) A typical graph of the log of the number of observed (infectious + removed) individuals as a function time. (b) The same function minus $0.05$ times the  time.}\label{traject}
\end{figure}
The simulations  are performed in R 
and in MATLAB. 
In all simulations we use a  Markov SEIR epidemic with the expected latent period twice the expected infectious period. This resembles the estimates for Ebola in West Africa \cite{teamebola},  where the average time between infection and symptom onset and the start of the infectious period is estimated to be approximately 9.4 days (standard deviation 7.4 days) and the average time between symptom onset and hospitalization or death is approximately 5 days (standard deviation 4.7 days). Because the differences between the means of the infectious and latent periods and their corresponding standard deviations are relatively small, we use a Markov SEIR epidemic model in which both periods are exponentially distributed. 

We simulated a Markov SEIR epidemic in a multi-type population 250 times in MATLAB. 
As a population we took the Dutch population in 1987  (approximately 14.6 million people) as used in \cite{Wall06}, for which extensive data on contact structure are available. The population is subdivided into six age groups (0-5, 6-12, 13-19, 20-39,40-59, 60+) and contact intensities are based on questionnaire data. For the simulations we use that the average infectious period $1/\gamma$ is 5 days, and the average latent period $1/\delta$ is 10 days.  The infection rates $\lambda_{ij}$ are chosen randomly for each simulation as follows.  The data in Table 1 of \cite{Wall06} give estimates of
$m_{ij}$ $(i,j=1,2,\dots,6)$, where $m_{ij}$ is the mean number of conversational partners per week in age class $i$ of a typical individual in age class $j$.  Using such conversations as a proxy for disease transmission, we assume that $\lambda_{ij}=c m_{ji} /\pi_j$,
where $\pi_j$ is the fraction of the Dutch population that are in age
class $j$, estimated from Appendix Table 1 in \cite{Wall06}, and $c$ is a multiplicative constant chosen so that $R_0^{(mult)}$ has a specified value, which is sampled independently and uniformly from the interval between 1.5 and 3 for each simulation.

All simulated epidemics start with 1 infectious individual in each of the six age groups. We use two estimates of $R_0$. The first of these estimates is based on the average number of offspring from the people who were infected as 100th up to 1000th. We ignore the first 100 infecteds to ignore the effect of the initial stages of the epidemic, when the proportions of infecteds are still far from equilibrium. This procedure leads to a very good estimate of $R_0$ if the spread of the disease is observed completely. 
The second estimate is based on $\hat{\alpha}$, an estimate of the epidemic growth rate $\alpha$, and neglects the multitype setting by assuming homogeneous mixing.
We assume that we know $\gamma$ and $\delta$ exactly and the estimate for $R_0$ is given by $(1+ \hat{\alpha}/\delta)(1+ \hat{\alpha}/\mu)$. The estimate $\hat{\alpha}$ is obtained from the development of the number of infectious people over time between the time the 100th  individual becomes infectious and the time the 1000th individual becomes infectious, by using least square estimation of the natural logarithm of the number of infecteds against time. More specifically, if $t_{100}, t_{101},\dots,t_{1000}$ denote the times that these individuals become infected then $\hat{\alpha}$ is obtained by fitting a straight line to the points $(\log(i),t_i), i=100,101,\dots,1000$ using linear regression, so
\[
\hat{\alpha}=
\frac{901 \sum_{i=100}^{1000} \log(i) t_i- \sum_{i=100}^{1000} \log(i) \sum_{j=100}^{1000} t_j}{901 \sum_{i=100}^{1000} t_i^2 - \left(\sum_{i=100}^{1000} t_i\right)^2}.
\]

In Figure 3(a)  we provide a scatter plot depicting the two estimates of $R_0$ for the 250 simulations. The ratio of the two estimates in the 250 simulations are summarized in Figure 3(b). We see that the estimates are generally very good, as predicted by the theory.

To simulate  epidemics on networks we use several networks from the Stanford Large Network Dataset collection \cite{Snap}. In Section 3 we use a collaboration network in Condense Matter physics, because $(i)$ this graph is undirected (if individual $a$ can contact individual $b$, then $b$ can contact $a$, $(ii)$ this graph is large (23133 individuals) and $(iii)$ the mean excess degree, $\kappa$ is not extremely high. Individuals are acquaintances if they were co-authors of a manuscript posted on the e-print service arXiv in the condense matter physics section between January 1993 and April 2004. A manuscript with more than 2 authors leads to cliques (small groups in which everybody is acquainted to everyone else in the group). Since arguably many networks relevant for the spread of infectious diseases contain such cliques (households, workplaces and groups of friends), the presence of many cliques in collaboration networks is a desirable property. 

Our simulations of Markov SEIR epidemics on all the  networks considered are performed in R, 
using the igraph package \cite{igraph}. An epidemic starts with 10 uniformly chosen individuals which are at the start of their infectious period at time 0. We estimate the epidemic growth rate $\alpha$ based on the time between the total number of individuals which are infectious or recovered/deceased (the individuals that have shown symptoms) increases from 200 to 400. We exclude all simulations in which the total number of affected individuals stays below 400.
The estimate of $R_0$ based on the real infection tree is obtained by looking at the epidemic from a generation perspective: All individuals infected by the initially infectious individuals are in generation 1, individuals infected by generation 1 infectives are in generation 2 etc.\ \cite{Pell12}. We consider as a reference generation the first generation in which there are 75 individuals (say generation $k$) and we divide the number of individuals in generation 2 up to $k+1$ by the number of individuals in generation 1 up to $k$. We exclude the initial individuals from the estimation of $R_0$, because those individuals are chosen uniformly at random and therefore independently of the population structure. 

By trial and error investigation we tune the infection parameter $\lambda$ such that the estimate of $R_0$ using the infection process is close to 2. Using this $\lambda$ we run 1000 simulations.
A typical graph of how the number of observed individuals (i.e.\ infectious + removed) is given in Figure \ref{traject}(a). In part $(b)$ we show the same graph but now we subtract $0.05$ times the time to show that the growth of the number of individuals is indeed close to exponential over a large time.

\begin{figure}[ht]

\centering

   \includegraphics[width=.8\textwidth]{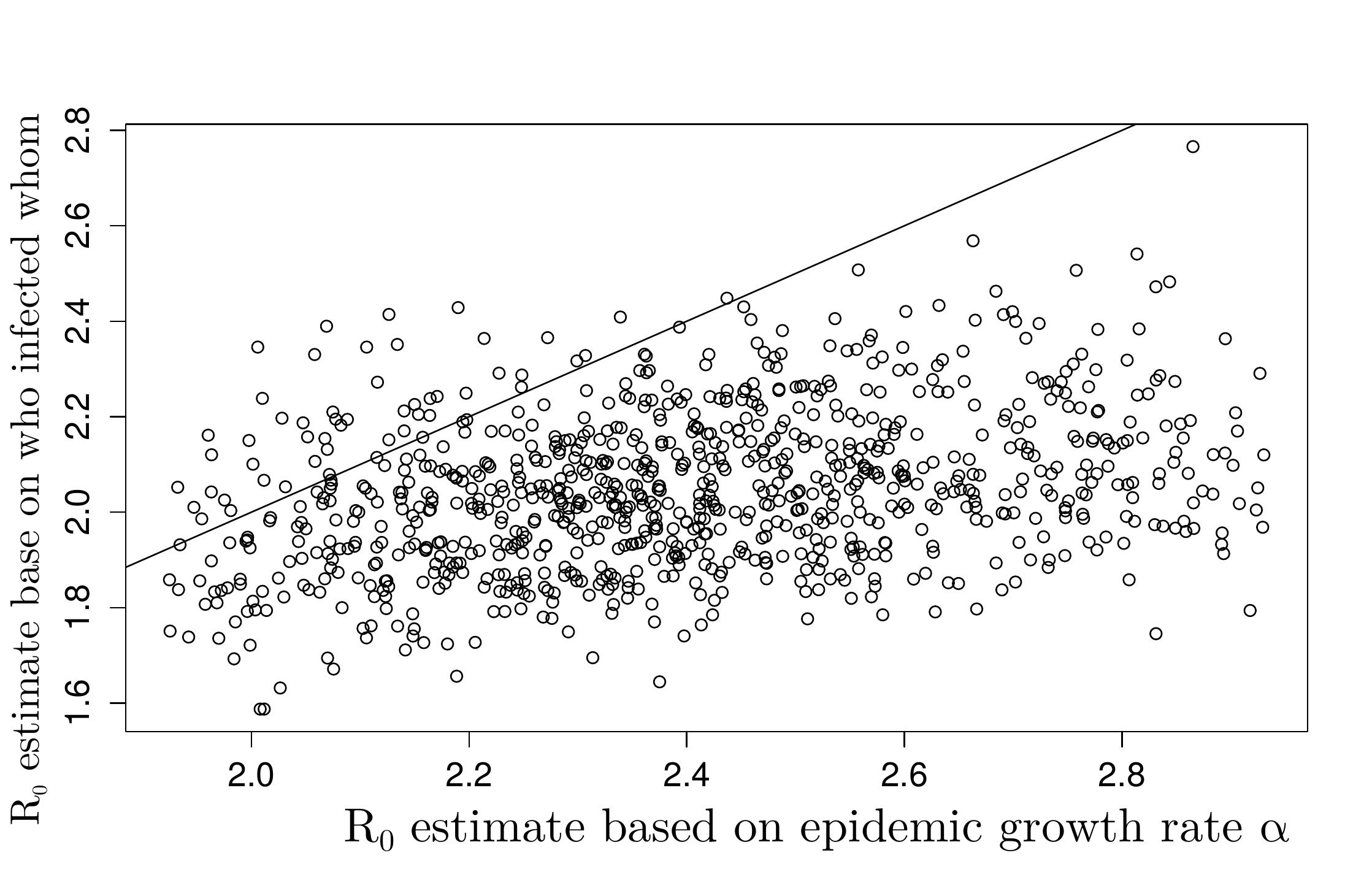}

\caption{Scatter plot of estimates of $R_0$ assuming homogeneous mixing and using the estimated epidemic growth rate, and estimates based on the real infection process (who infected whom) in the collaboration network in Condense Matter Physics. 1000 simulations are used and the simulations with the 50 lowest and 50 highest estimated epidemic growth rates are not represented in the scatter plot. The line shows where the two estimates are equal.}\label{scattercond}
\end{figure}

Because of the mechanical way of estimating $\alpha$, it is possible to have atypical epidemic trajectories, in which the estimation procedure is not good. Examples are (i) epidemics in which for example the exponential growth has not started yet at the time the 200th individual starts its infectious period or (ii) epidemics
where just around the time the  200th or 400th of individual starts its infectious period a new part of the network is affected, where this new part contains many acquaintances within itself but is not well connected to the rest of the network. Such an event causes a sudden strong increase in the observed cases. 
These atypical trajectories are possible to identify if one observes the number of infectious individuals for a single epidemic and better estimates can be obtained in this way.
We deal with this problem by not considering the simulations which give the $5\%$ lowest and $5\%$ highest estimates for $\alpha$.

In Figure \ref{scattercond} we provide a scatter plot of the two estimates of $R_0$ for the simulations used, we see that in the vast majority of the simulations, the estimate of $R_0$ based on the estimated $\alpha$ and the homogeneously mixing assumption is conservative. We note that the two estimates are hardly correlated. 

\begin{figure}[ht]

\centering
\subfigure[ ]{
    \includegraphics[width=.45\textwidth]{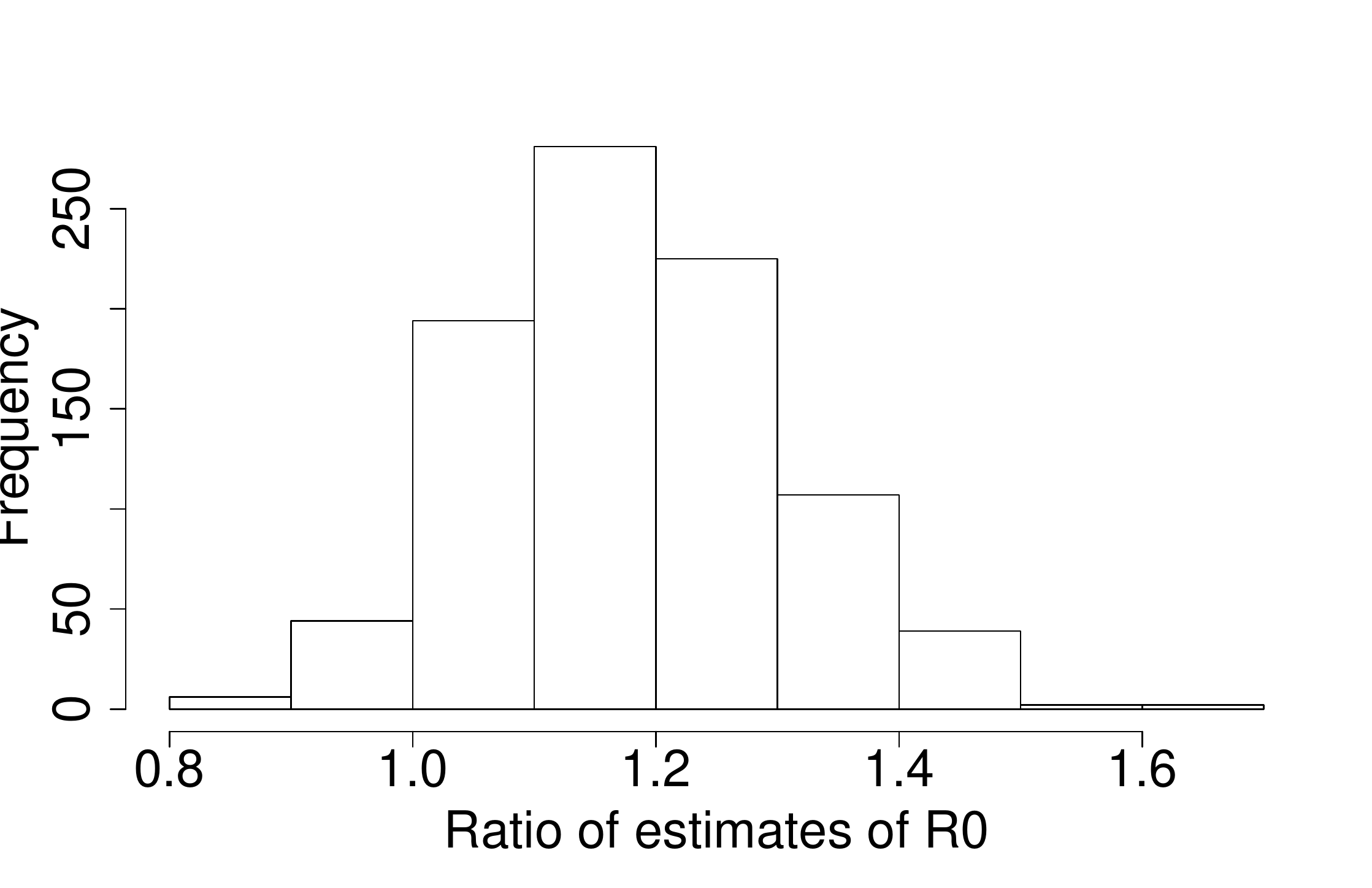}
}
\subfigure[ ]{
   \includegraphics[width=.45\textwidth]{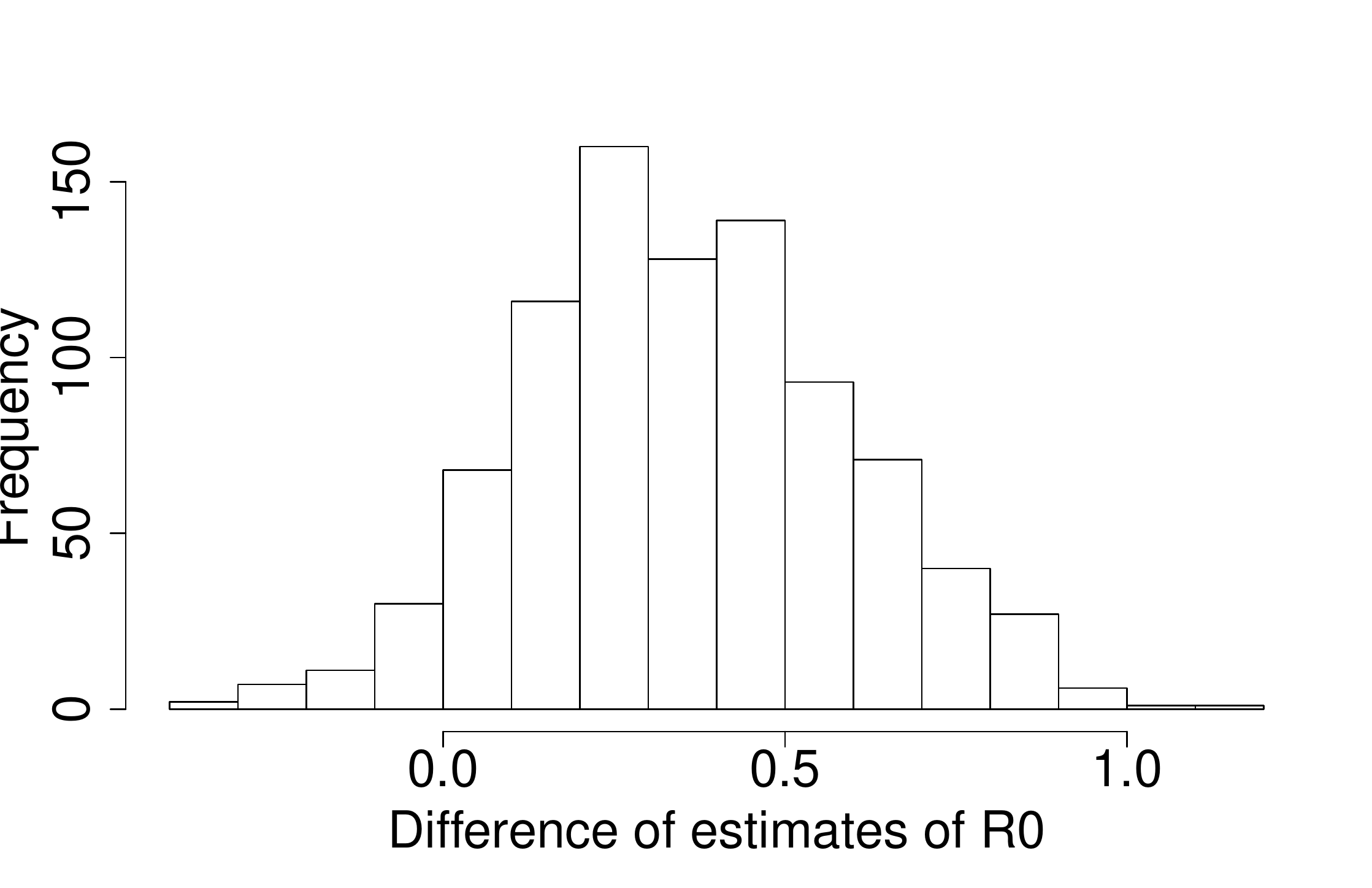}
}
\caption{Histograms of the ratio (a) of and difference (b) between the estimates of $R_0$ assuming homogeneous mixing and using the estimated epidemic growth rate, and estimates based on the real infection process in the collaboration network in Condense Matter Physics. 1000 simulations are used and the simulations with the 50 lowest and 50 highest estimated epidemic growth rates are not represented in the histograms.}\label{histscond}
\end{figure}

We further summarize our data in Figure 5, and in Figure \ref{histscond}. In which the ratio and difference of the $R_0$ estimate based on the epidemic growth rate and assuming homogeneous mixing, and the $R_0$ estimate based on the observed infection process, are given.

\begin{figure}[ht]

\centering
\subfigure[ ]{
\includegraphics[width=.6\textwidth]{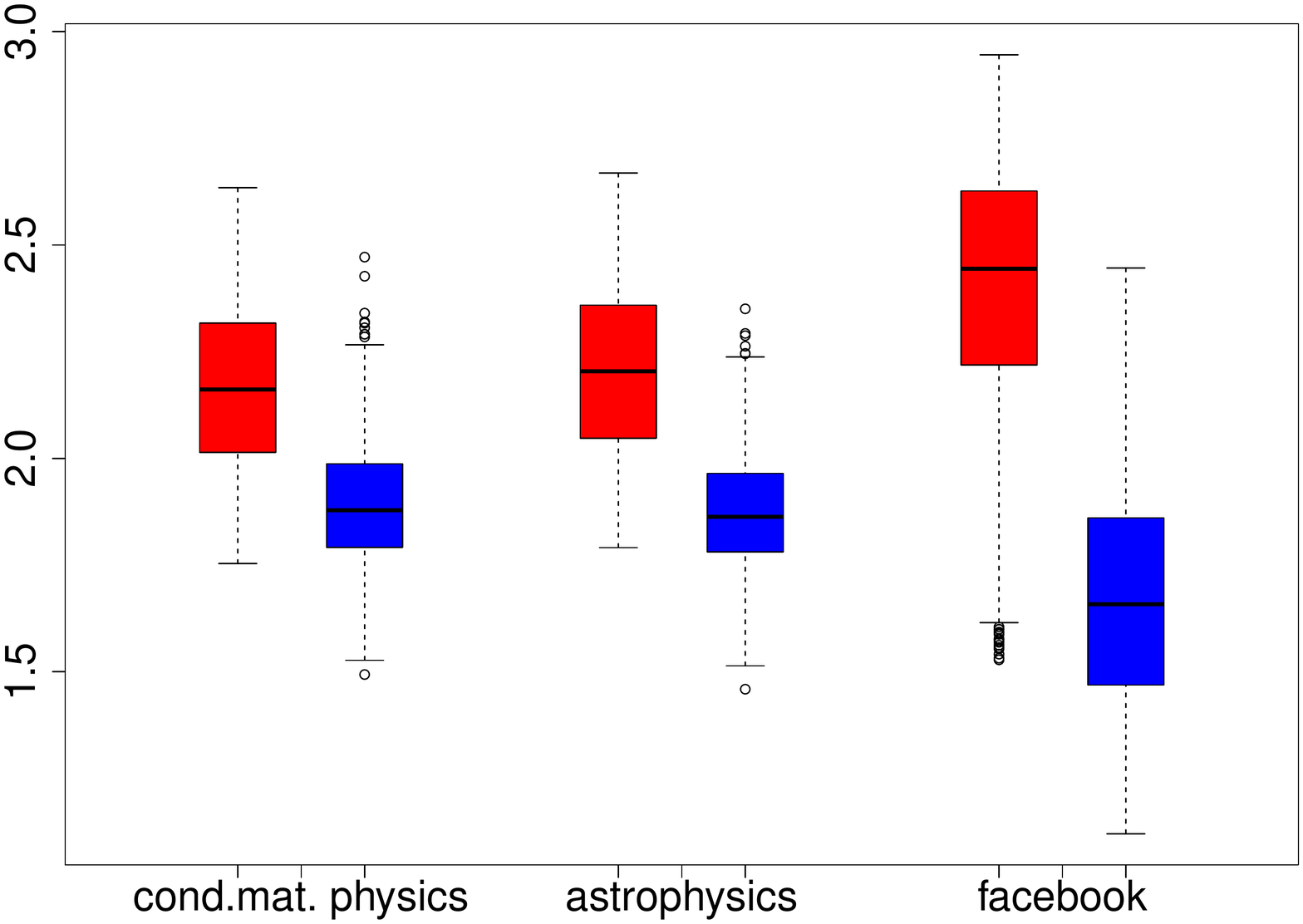}}
\subfigure[ ]{
\includegraphics[width=.35\textwidth]{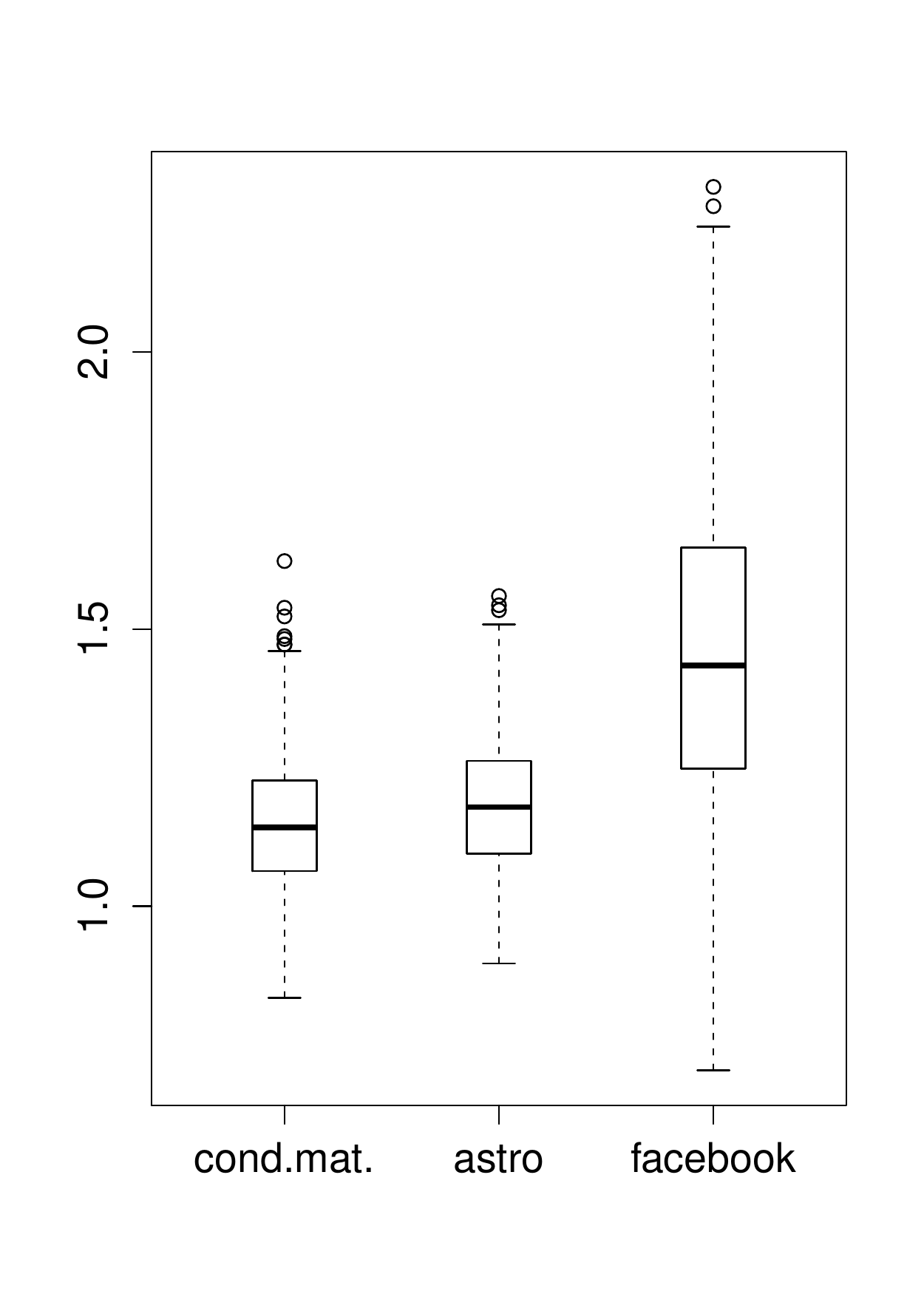}
}
\caption{Boxplots of estimates of $R_0$ for three networks from \cite{Snap}: The condensed matter physics and astrophysics collaboration network and a facebook social network graph. In (a) the estimates assuming homogeneous mixing and using the epidemic growth rate are plotted in red, while the estimates based on the real infection process are plotted in blue. In (b) the ratios of the two estimates of $R_0$ for each simulation are summarized.}\label{boxplots}
\end{figure}

We also analyse the spread of SEIR epidemics on 2 other networks described in the Stanford Large Network Dataset collection \cite{Snap}. The first is the collaboration network in Astro Physics, which is obtained in a similar way as the collaboration network in Condense Matter Physics. This network is slightly smaller than the Condense Matter Physics network and has a higher $\kappa$ (approximately 64 instead of 21). The analysis is performed similarly to the analysis of the Condense Matter Physics collaboration network. Boxplots of the estimates of $R_0$ using the real infection process, the estimates of $R_0$ using the epidemic growth rate and assuming homogeneous mixing, as well as a boxplot of the ratio of those estimates, are given in Figure \ref{boxplots}.

We see that the two estimates are close, but that the simpler estimate assuming homogeneous mixing is slightly conservative for all three empirical networks, which is consistent with the theoretical result for the configuration model.   

The second alternative network is a part of the facebook social network from \cite{Snap}. This part is relatively small and we restrict ourselves to the largest connected component (containing 1034 individuals). This network has a high mean degree (51.7) and mean excess degree (93.5). Because of its relatively small size, and the observation that some substantial parts of the network are connected to the other parts of the network through only a few connections, the estimate of $R_0$ through the epidemic growth rate is less good. We also have to adapt the bounds for estimating $R_0$ from the infection tree (as a reference generation the first generation in which there are 40 individuals), and we estimate the epidemic growth rate based on the time between the total number of individuals which are infectious or recovered/deceased  increases from 150 to 350. Furthermore, in order to obtain quicker convergence the 7 initial infectious individuals are chosen proportional to their number of acquaintances, which gives individuals with many acquaintances a higher probability of being initially infectious. 
Boxplots of the estimates of $R_0$ using the real infection process, the estimates assuming homogeneous mixing and using the epidemic growth rate, as well as a boxplot of the ratio of those estimates, are given in Figure \ref{boxplots}.

\newpage

\bibliographystyle{siam}
\bibliography{publications.bib}
\end{document}